%% file: AA_main_file.tex
\documentclass[fleqn,usenatbib]{mnras}
\pdfoutput=1
\usepackage{ulem} 

\usepackage[T1]{fontenc}

\DeclareRobustCommand{\VAN}[3]{#2}
\let\VANthebibliography\thebibliography
\def\thebibliography{\DeclareRobustCommand{\VAN}[3]{##3}\VANthebibliography}

\usepackage{graphicx}	
\usepackage{amsmath}	
\usepackage{amssymb}	
\usepackage{hyperref}
\usepackage{newtxtext,newtxmath}

\title[Four New Deeply-Eclipsing WDs]{Four New Deeply-Eclipsing White Dwarfs in ZTF}

\author[Kosakowski et al.]{
A. Kosakowski,$^{1,2}$
M. Kilic,$^{2}$
W. R. Brown,$^{3}$
P. Bergeron,$^{4}$
and
T. Kupfer$^{1}$
\\
$^{1}$Department of Physics and Astronomy, Texas Tech University, Lubbock, TX 79409, USA\\
$^{2}$Homer L. Dodge Department of Physics and Astronomy, University of Oklahoma, 440 W. Brooks St., Norman, OK, 73019 USA\\
$^{3}$Smithsonian Astrophysical Observatory, 60 Garden St, Cambridge, MA 02138 USA \\
$^{4}$Département de Physique, Université de Montréal, C.P. 6128, Succ. Centre-Ville, Montréal, Québec H3C 3J7, Canada \\
}

\date{Accepted XXX. Received YYY; in original form ZZZ}

\pubyear{2022}

\begin{document}
\label{firstpage}
\pagerange{\pageref{firstpage}--\pageref{lastpage}}
\maketitle

\begin{abstract}
We present the results of a search for deeply-eclipsing white dwarfs in the ZTF Data Release 4. We identify nine deeply-eclipsing white dwarf candidates, four of which we followed up with high-cadence photometry and spectroscopy. Three of these systems show total eclipses in the ZTF data and our follow-up APO 3.5-meter telescope observations. Even though the eclipse duration is consistent with sub-stellar companions, our analysis shows that all four systems contain a white dwarf with low-mass stellar companions of $\sim0.1~M_{\odot}$. We provide mass and radius constraints for both stars in each system based on our photometric and spectroscopic fitting. Finally, we present a list of 41 additional eclipsing WD+M candidates identified in a preliminary search of ZTF DR7, including 12 previously studied systems. We identify two new candidate short-period, eclipsing, white dwarf-brown dwarf binaries within our sample of 41 WD+M candidates based on PanSTARRS colors.
\end{abstract}




\section{Introduction}

Because white dwarfs are compact objects with typical radii $\sim0.01~R_{\odot}$, their stellar and sub-stellar companions can cause significant eclipses. Such eclipses can be identified in large scale photometric surveys that are deep enough to detect a significant number of white dwarfs. For example, \citet{vanderburg2015,vanderburg2020} used data from the Kepler K2 mission and the Transiting Exoplanet Survey Satellite (TESS) to discover $\sim50$\% eclipses due to a disintegrating minor body around WD 1145$+$017 and grazing eclipses of a candidate giant planet around WD 1856$+$534. Similarly, \citet{parsons2017} used Kepler K2 data to identify two eclipsing binaries containing white dwarfs with cool companions8, including the total eclipse of the white dwarf + brown dwarf system SDSS J1205$-$0242.

The Zwicky Transient Facility \citep[ZTF;][]{bellm2019,masci2019}, in combination with the large number of white dwarfs identified through Gaia parallaxes \citep{gentilefusillo2019,gentilefusillo2021}, provides an excellent opportunity to identify variable white dwarfs \citep[e.g.,][]{vanderbosch2020,guidry2021}. \citet{burdge2020a,burdge2020b,coughlin2020} and \citet{vanroestel2021a} used ZTF data to identify 22 binary systems with orbital periods less than an hour, including nine eclipsing detached binaries. Eclipse depths range from a few percent to almost 100\% in eclipsing double white dwarf systems currently known \citep[e.g.,][]{steinfadt2010,brown2011,kilic2014,hallakoun2016,brown2017,burdge2020a}. \citet{vanroestel2021b} expanded this search to deeply-eclipsing white dwarfs and identified the 10 hour period eclipsing white dwarf + brown dwarf ZTF J0038$+$2030. 

There are $\approx10$ short period, detached white dwarf + brown dwarf binaries known \citep[][and references therein]{hogg2020}, but only four are eclipsing \citep{beuermann2013,parsons2017,casewell2020,vanroestel2021b}. The deeply-eclipsing white dwarfs with stellar or sub-stellar companions offer a unique opportunity to obtain high precision mass and radius measurements of both components of the binary. These constraints are especially valuable for unusual systems, or systems where there are large discrepancies between the theoretical and observed radii measurements, such as low-mass M dwarfs \citep[e.g.,][]{lopez-morales2005}. \citet{parsons2018} obtained model-independent masses and radii for 23 M dwarfs in binaries with white dwarfs, and found a large scatter in the M dwarf radii, with 75 percent of their objects being up to 12 percent over-inflated. \citet{kesseli2018} argue that all fully convective M dwarfs are larger than model predictions by 13-18\% for the lowest mass, $0.08~{\rm M_\odot}< M < 0.18~{\rm M_\odot}$ M dwarfs.

Here we present the results from a search for deeply-eclipsing white dwarfs in the ZTF Data Release 4. We identify nine deeply-eclipsing white dwarfs and obtain follow-up spectroscopy and photometry of four of these systems with periods less than $P\approx3~{\rm h}$. We show that all four systems with follow-up observations contain low-mass M dwarf companions with $M\sim0.1~{\rm M_\odot}$. We discuss our target selection in Section 2, our observations and analysis of each object in Section 3, and discuss their fitted parameters and conclude in Section 4.

\input{figures/4_ztf_lightcurves.tex}

\section{Target Selection}

\citet{gentilefusillo2019} identified 486,641 white dwarf candidates using Gaia Data Release 2 photometry and astrometry. Their catalogue is up to 85\% complete for white dwarfs brighter than ${\rm G} = 20~{\rm mag}$ and with $T_{\rm eff}>7000~{\rm K}$ at Galactic latitudes $|b|>20^\circ$. We queried the ZTF Data Release 4 public data archive for all light curves within $5\arcsec$ of each of these 486,641 targets. Our relatively large search radius ensures that high proper motion objects are not lost in the ZTF search. We exclude objects with fewer than 27 epochs and remove photometry flagged as bad in ZTF. Because the ZTF astrometry algorithm assigns different object IDs to the same object in different filters, and occasionally the same object in the same filter, we combine data for objects within $2.5\arcsec$ of one another and normalize their magnitudes to their median ZTF $g$-band magnitude, or their median ZTF $r$-band magnitude if the ZTF $g$-band data is not available. By combining the light curves across multiple filters, we increased the temporal sampling for each light curve and improve our ability to detect variability in each light curve.

Our query returned a total of 230,870 combined light curves, some of which are of nearby field stars, unrelated to the white dwarfs in our target list, due to our relatively large search radius. Finally, because we are specifically searching for deeply-eclipsing systems, we require that each light curve has at least seven $4\sigma$ or five $5\sigma$ deviant points from the its quiescent level, defined as the object’s light curve after performing three iterations of $3\sigma$ clipping around its median magnitude value. Our final sample included 2498 light curves with magnitudes between $12.8~{\rm mag}$ and $20.8~{\rm mag}$ (median $18.6~{\rm mag}$) and between 88 and 5740 epochs (median 927 epochs).

\section{Period Search Results}

We used a box least squares \citep[BLS;][]{kovacs2002} period finding algorithm to search for deeply-eclipsing white dwarf binaries with periods between $P_{\rm min}=5~{\rm min}$ and $P_{\rm max}=11.9~{\rm h}$ within our sample of 2498 light curves from the ZTF Data Release 4 (DR4) public data archive. Because the BLS algorithm favors sharp ingress and egress features, it is ideally suited for finding deeply-eclipsing white dwarfs.
Our BLS algorithm assigned 200 bins to each light curve and fit a box-shaped eclipse with total duration equal to $0.1\%$ to $12.5\%$ of the orbital period.
Our BLS search identified nine such binaries. We present our follow-up analysis to four of these binaries with periods less than $P=3~{\rm h}$: ZTF J164441.1946$+$243428.2112 (J1644$+$2434), ZTF J174424.7141$+$390215.6653 (J1744$+$3902), ZTF J184434.3978$+$485736.5063 (J1844$+$4857), and ZTF J221226.9672$+$534750.6967 (J2212$+$5347).

Figure \ref{fig:4_ztf_lightcurves} shows the phase-folded ZTF DR4 light curves of these four systems with periods ranging from $1.4~{\rm h}$ to $2.7~{\rm h}$. All four systems show $\sim5$ min long eclipses, consistent with low-mass M dwarf, brown dwarf, or massive giant planet companions (see \citet{rappaport2021} for a discussion on the minimum allowed orbital period of H-rich bodies of varying mass).
The eclipse bottom is not detected for any of the targets in the ZTF $g$-band data, but it is detected in the $r$- and $i$-bands for J1844$+$4857. The eclipse depth is much shallower in the $i$-band for J1844$+$4857, indicating that the companion contributes significant flux in the $i$-band. J1744$+$3902 shows a significant reflection effect, confirming that it has a relatively cool companion. 

Three of these systems, J1644$+$2434, J1844$+$4857, and J2212$+$5347, were independently identified as eclipsing white dwarfs by \citet{keller2022} in a similar box least squares search for short-period white dwarf binaries in the ZTF DR3 archive. However, the authors did not provide any follow-up observations or detailed light curve analysis of these systems.

\section{Follow-up Observations and Data Analysis}

\subsection{Spectroscopy}

\input{figures/4_spectra.tex}

We used the Apache Point Observatory 3.5-meter telescope's Dual Imaging Spectrograph (DIS) with the B400 + R300 gratings and a $1.5\arcsec$ slit to obtain spectroscopy of J1644$+$2434 and J1744$+$3902 on UT 2021 April 09. The blue and red channels provide spectral resolutions of 5.2 and 6.4 \AA\ over the wavelength ranges $3400 - 5500$ and $5500 - 9200$ \AA, respectively. We obtained a single exposure of 1600-1800 seconds for each object.

We obtained time-series spectroscopy of J1844$+$4857 and J2212$+$5347 using the 6.5-meter MMT with the Blue Channel Spectrograph. We operated the spectrograph with the $832~{\rm lines~mm^{-1}}$ grating in second order and a $1.25\arcsec$ slit, providing wavelength coverage from 3600 \AA\ to 4500 \AA\ and a spectral resolution of 1.25 \AA. We obtained eight back-to-back spectra of J1844$+$4857 on UT 2021 June 16 with five additional spectra obtained on June 18. For J2212$+$5347, we obtained 27 spectra spread across four nights on UT 2020 December 9, 12, 15, and 16.

Figure \ref{fig:4_spectra} shows the APO and MMT spectra of our four deeply-eclipsing white dwarfs. All four systems are confirmed to contain DA white dwarfs, with no obvious features from their companions in their blue spectra. J2212$+$5347 is the only object to show a photospheric Ca K absorption line.

Because the cool companions do not contribute significantly to the blue spectrum of these systems, we performed spectroscopic fitting to the blue-optical data using a grid of pure-hydrogen white dwarf model atmospheres and obtain best-fitting atmospheric parameters $T_{\rm eff}$ and $\log{g}$ for each white dwarf. The details of our fitting procedure are described in \cite{gianninas2014}. Our best-fitting models are over-plotted onto the Balmer lines of our observed spectra in Figure \ref{fig:4_spectra_fits}. We add in quadrature the external uncertainties of $\sigma_{T_{\rm eff}}\approx1.2\%$ and $\sigma_{\log{g}}\approx0.038~{\rm dex}$ from \citet{liebert2005} to our presented results.

\input{figures/4_spectra_fits.tex}

We obtained radial velocity measurements for each of our spectra for J1844$+$4857 and J2212$+$5347 using the cross-correlation fitting package \textsc{rvsao} \citep{kurtz1998} within \textsc{iraf}. We created a zero-velocity summed spectrum for each object and use these high signal-to-noise summed spectra as templates for our cross-correlation. We applied barycentric correction to the resulting radial velocity measurements of each individual spectrum. Our radial velocity measurements for both J1844$+$4857 and J2212$+$5347 are presented in Table \ref{tab:rv_table_1844_2212}. We fit a circular orbit to our radial velocity measurements for each object using a Monte Carlo approach based on \citet{kenyon1986} to obtain measurements of the velocity semi-amplitude for each system.

\subsection{High-cadence Photometry}

We obtained follow-up high-speed photometry for all four targets using the Apache Point Observatory 3.5-meter telescope's frame transfer camera, Agile \citep{mukadam2011}. We binned the CCD by $2\times2$, which resulted in a plate scale of $0.258~{\rm arcsec~pixel^{-1}}$. We used the ZTF light curves to predict the eclipse times for each system during our observing runs and observed them during those windows to acquire data right before, during, and after an eclipse. We obtained broadband BG40 photometry of J1644$+$2434 on UT 2021 April 3 and 4, covering an eclipse on each night. Similarly, we obtained SDSS $r$-band photometry of J1744$+$3902 on UT 2021 April 3 and 4 and $i$-band photometry on UT 2021 April 9, covering an eclipse in each filter/night. Given the relatively short orbital period of 1.7 h for J1844$+$4857, we were able to observe two consecutive eclipses in the BG40 filter on UT 2021 April 4. Finally, for our shortest period system, J2212$+$5347, we obtained 380 minutes of continuous BG40 filter photometry on UT 2020 December 7.

We used \textsc{iraf} \citep{tody1986} to perform forced photometry using a varying aperture size based on the average ${\rm PSF}$ size of each image. We used relatively bright sources in the Agile field-of-view to calibrate the photometry for each source. We performed light curve fitting to our follow-up APO light curves using JKTEBOP \citep{southworth2013}.

For J1644$+$2434 and J1744$+$3902, which do not have radial velocity constraints, we fit for the orbital inclination and the sum and ratio of the fractional component radii, defined as the stellar radii divided by the orbital separation ($r=R/a$). We used gravity darkening and quadratic limb darkening coefficients from \cite{claret2020} for DA white dwarfs based on the atmospheric $T_{\rm eff}$ and $\log{g}$ values obtained from our spectroscopic fits. For the companion, we used a gravity darkening coefficient from \citet{claret2017} and linear limb darkening coefficients of \citet{claret2012} for a $T_{\rm eff}=3000~{\rm K}$ $\log{g}=5.00$ low mass stellar object. We allow the limb darkening coefficients to vary within $\approx10\%$ of their initial values to account for the uncertainty in our stellar parameters. We converted the $ugr$-band coefficients into BG40 coefficients using equation 3 in \cite{hallakoun2016} when fitting our BG40 light curves. We performed 10,000 Monte Carlo fits to the light curves with initial parameters based on typical low-mass stellar radii values combined with our white dwarf parameters obtained through spectroscopy.

For J1844$+$4857 and J2212$+$5347, we make use of our radial velocity constraints and follow the methods of \citet{parsons2017} to estimate the component radii, orbital separation, orbital inclination, and companion mass. In short: we performed Monte Carlo fits to our light curves using JKTEBOP with fixed inclination, ranging from $75^\circ$ to $90^\circ$ in steps of $1^\circ$, fitting for the fractional component radii. We then used our radial velocity constraints to estimate the companion mass at each fitted inclination and compared our results with the low mass stellar models of \citet{baraffe2015} for various system ages. In each case, the low mass stellar models highlight a single inclination as the most-probable inclination of the binary. Finally, we performed 10,000 additional Monte Carlo fits for the stellar fractional component radii and orbital inclination, allowing the inclination to vary from this most-probable initial value. We used the quadratic limb darkening coefficients from \cite{claret2020} for the white dwarf primary and the linear limb darkening coefficients from \citet{claret2017} for the low-mass stellar companion. We allow our limb darkening coefficients to vary within $\approx10\%$ of their initial values to account for the uncertainties in our stellar parameters, with the exception of the linear coefficient to the companion of J2212$+$5347 due to our fits failing to converge on physical solutions with variable companion limb darkening enabled. Through this method, we estimated the mass and radius of the low-mass stellar companions, based on the \citet{baraffe2015} models, our radial velocity measurements, and our eclipse fitting.

\section{Binary Parameters and Results}
\subsection{ZTF J164441.1946$+$243428.2112}

J1644$+$2434 is a DA white dwarf in a binary with a low-mass stellar companion with orbital period $P=115.35~{\rm min}$. Spectroscopic fits suggest atmospheric parameters $T_{\rm eff}=14,900\pm760~{\rm K}$ and $\log{g}=7.89\pm0.14$ for the primary star, corresponding to a white dwarf with $M_{\rm WD}=0.55\pm0.07~{\rm M_\odot}$ and $R_{\rm WD}=0.014\pm0.001~{\rm R_\odot}$ based on the cooling sequences\footnote{https://www.astro.umontreal.ca/$\sim$bergeron/CoolingModels/} of \citet{bedard2020}.

Our fits to the BG40 broadband light curve return most-probable parameters $r_B+r_A=0.201^{+0.008}_{-0.007}$, $\frac{r_B}{r_A}=4.2^{+0.7}_{-0.6}$, and $i=83.6\pm0.9^\circ$. We present our resulting parameter distributions with their most-probable values and $1\sigma$ uncertainties in Figure \ref{fig:1644_apo_bg40_fit}. Combined with the model-dependent white dwarf radius obtained through spectroscopic fits, we calculated the companion radius $R_2=0.06\pm0.01~{\rm R_\odot}$ and orbital separation $a=0.36\pm0.07~{\rm R_\odot}$. However, since we do not detect the eclipse minimum, our parameter estimates act as limits to their true values, which depend strongly on both the depth of eclipse and the shape of the ingress and egress features. An increase in inclination would result in a corresponding decrease in companion radius or increase in orbital separation, $a$. Given the unusually small companion radius compared to low-mass stellar models, it is likely that our fitted inclination is an upper limit while the companion fractional radius is a lower limit. A deeper eclipsing light curve is required to obtain more precise estimates of the stellar parameters in this binary.

We estimated the mass of the companion by fitting the available PanSTARRS SED with a composite model containing a pure-hydrogen white dwarf with parameters based on our spectroscopic fit and the low-mass stellar object models of \citet{baraffe2015}. Our best-fitting companion mass is $M_{\rm 2,SED}=0.084^{+0.004}_{-0.003}~{\rm M_\odot}$, placing the companion near the Hydrogen-burning limit at $M=73\sim81~{\rm M_{Jupiter}}$. We present our best-fit composite SED model for J1644$+$2434 in Figure \ref{fig:2_sed_models} (top).

\input{figures/1644_apo_bg40_fit.tex}
\input{figures/2_sed_models.tex}

\subsection{ZTF J174424.7141$+$390215.6653}

J1744$+$3902 is a nearby ($d_\pi=241\pm5~{\rm pc}$) eclipsing binary containing a DA white dwarf and a low-mass stellar companion with orbital period $P=161.97~{\rm min}$. Our fits to the blue-optical spectrum provide white dwarf atmospheric parameters $T_{\rm eff}=13,830\pm360~{\rm K}$ and $\log{g}=7.61\pm0.07$, corresponding to a white dwarf with $M_{\rm WD}=0.41\pm0.03~{\rm M_\odot}$ and $R_{\rm WD}=0.017\pm0.001~{\rm R_\odot}$ based on the cooling sequences of \citet{bedard2020} or $M_{\rm WD,He}=0.43\pm0.03~{\rm M_\odot}$ and $R_{\rm WD,He}=0.017~{\rm R_\odot}$ based on He-core models of \citet{althaus2013}.

We obtained $r$- and $i$-band light curves of J1744$+$3902. Our $r$-band light curve contains two primary eclipses obtained on separate nights, while our $i$-band light curve contains only a single primary eclipse. The most-probable fitted parameters for each filter agree within one sigma. However, our $i$-band light curve is poorly-sampled during the ingress and egress when compared to our $r$-band light curve. We therefore use the $r$-band values when reporting the most-probable parameters and their uncertainties. The most-probable fitted parameters for J1744$+$3902 are $r_B+r_A=0.218\pm0.002$, $\frac{r_B}{r_A}=8.2\pm0.2$, and $i={87.2^{+0.6}_{-0.5}}^\circ$, corresponding to a companion radius $R_2=0.14\pm0.01~{\rm R_\odot}$, in agreement with the low-mass stellar models of \citet{baraffe2015}. We present these parameter distributions and light curves for the $r$- and $i$-band data of J1744$+$3902 in Figures \ref{fig:1744_apo_rband_fit2} and \ref{fig:1744_apo_iband_fit2}, respectively. A higher signal-to-noise mid-eclipse light curve with improved phase sampling is desired in order to better constrain the companion's physical parameters.

\input{figures/1744_apo_rband_fit2}
\input{figures/1744_apo_iband_fit2}

We used the available PanSTARRS SED to estimate the mass of the companion to be $M_2=0.121^{+0.006}_{-0.007}~{\rm M_\odot}$, well above the Hydrogen burning limit and in good agreement with the radius estimate for a $\tau=5~{\rm Gyr}$, $R=0.141\pm0.008~{\rm R_\odot}$ low-mass stellar object estimated from our eclipsing light curve fits. Our SED fit to J1744$+$3902 is presented in Figure \ref{fig:2_sed_models} (bottom).

\subsection{ZTF J184434.3978$+$485736.5063}
J1844$+$4857 is a DA white dwarf in an eclipsing binary with a low-mass stellar object with orbital period $P=104.64~{\rm min}$. Our spectroscopic fits to the blue-optical spectrum return best-fitting atmosphere parameters $T_{\rm eff}=18,040\pm320~{\rm K}$ and $\log{g}=7.61\pm0.06$, corresponding to a white dwarf with $M_{\rm WD}=0.43\pm0.02~{\rm M_\odot}$ and $R_{\rm WD}=0.017\pm0.001~{\rm R_\odot}$ based on the cooling sequences of \citet{bedard2020} or $M_{\rm WD,He}=0.45\pm0.03~{\rm M_\odot}$ and $R_{\rm WD,He}=0.017~{\rm R_\odot}$ based on He-core models of \citet{althaus2013}. 

\input{figures/1844_mmt_rv_figure.tex}

We fit a circular orbit to the 13 radial velocity measurements of J1844$+$4857. Due to the large radial velocity uncertainties caused by the binary's relative faintness ($G=19.3~{\rm mag}$), we fixed the orbital period at the photometric value $P=1.744~{\rm h}$, obtained from a baseline of over 800 days of ZTF observations. Our best-fitting model returns systemic velocity $\gamma=-38\pm14~{\rm km~s^{-1}}$ and velocity semi-amplitude $K=109\pm23~{\rm km~s^{-1}}$. We present the fitted radial velocities for J1844$+$4857 in Figure \ref{fig:1844_mmt_rv_figure}. We used the binary mass function to estimate the minimum companion mass $M_{2,{\rm min}}=155\pm40~{\rm M_{Jupiter}}$, assuming inclination $i=90^\circ$. Despite the relatively large uncertainties, the companion mass for J1844$+$4857 is clearly well above the upper mass limit for sub-stellar objects of $M\approx73~{\rm M_{Jupiter}}$ \citep{chabrier1997}.

Our eclipsing light curve fits to J1844$+$4857 suggest a primary white dwarf radius $R_{\rm WD}=0.0167^{+0.0009}_{-0.0008}~{\rm R_\odot}$, companion radius $R_B=0.184^{+0.0004}_{-0.0005}~{\rm R_\odot}$, and orbital inclination $i=76.3\pm0.5^\circ$, corresponding to a companion mass $M_2=0.16\pm0.04~{\rm M_\odot}$, with uncertainties dominated by our relatively large uncertainty in the radial velocity semi-amplitude. Figure \ref{fig:1844_parsons2017_fig6} displays our fitted parameter values on a companion mass-radius plot using the low-mass stellar models of \citet{baraffe2015}. These models suggest a true inclination of $i_0\approx76^\circ$. Our companion radius and orbital inclination both agree within $1\sigma$ of the low-mass stellar models, while our white dwarf radius estimate agrees within $1\sigma$ of our spectroscopic value. We present our fitted parameter distributions and light curve in Figure \ref{fig:1844_apo_bg40_fit} with most-probable parameter values and their $1\sigma$ ranges printed above each 1D distribution.

We fit the available PanSTARRS SED of J1844$+$4857 to estimate its companion mass and confirm the mass obtained through our combined radial velocity and eclipsing light curve fitting. Our best-fitting composite model returns $M_2=0.137^{+0.010}_{-0.009}~{\rm M_\odot}$ as the mass of the companion, in agreement with our estimate from radial velocity and eclipsing light curve fitting.

\input{figures/1844_parsons2017_fig6.tex}
\input{figures/1844_apo_bg40_fit.tex}

\input{tables/rv_table_1844_2212.tex}

\subsection{ZTF J221226.9672$+$534750.6967}

J2212$+$5347 is a nearby ($d_\pi=195\pm4~{\rm pc}$) eclipsing binary containing a low-mass DA white dwarf and a low-mass stellar companion in a compact orbit with orbital period $P=84.01~{\rm min}$. Our spectroscopic fits to the blue-optical spectrum suggest a primary white dwarf with atmospheric parameters $T_{\rm eff,0}=9560\pm120~{\rm K}$ and $\log{g}_0=7.86\pm0.04$. However, because cool white dwarfs return systematically large surface gravity when fit with 1D models \citep[see][]{tremblay2011}, we applied a 3D correction to our fitted values based on the correction equations provided in \citet{tremblay2015}, resulting in 3D-corrected values $T_{\rm eff}=9430\pm120~{\rm K}$ and $\log{g}=7.56\pm0.04$, corresponding to a white dwarf mass $M_{\rm WD}=0.38\pm0.02~{\rm M_\odot}$ and radius $R_{\rm WD}=0.0169\pm0.0004$ based on the white dwarf cooling sequences of \citet{bedard2020}, or $M_{\rm WD,He}=0.40\pm0.02~{\rm M_\odot}$ and $R_{\rm He}=0.017~{\rm R_\odot}$ based on the He-core white dwarf models of \citet{althaus2013}. Our best-fitting He-core solution suggests a distance $d_{\rm He}=196~{\rm pc}$ based on apparent magnitude and temperature measurements, in nearly perfect agreement with the distance based on Gaia eDR3 parallax measurements $d_\pi=195\pm4~{\rm pc}$.

\input{figures/2212_mmt_rv_figure.tex}

We fit a circular orbit to the 27 radial velocity measurements of J2212$+$5347 and obtained best-fitting orbital parameters $\gamma=-33\pm3~{\rm km~s^{-1}}$, $K=82.5\pm4~{\rm km~s^{-1}}$, and $P=1.4000\pm0.0001~{\rm h}$, corresponding to a minimum companion mass $M_{2,{\rm min}}=95\pm6~{\rm M_{Jupiter}}$ ($0.091\pm0.006~{\rm M_\odot}$); well above the mass limit for sub-stellar objects. Our best-fitting circular orbit for J2212$+$5347 is presented in Figure \ref{fig:2212_mmt_rv_figure}. Our radial velocity measurements are presented in Table \ref{tab:rv_table_1844_2212}.

\input{figures/2212_parsons2017_fig6.tex}
\input{figures/2212_apo_bg40_fit.tex}

Our eclipsing light curve fits to J2212$+$5347 suggest a primary white dwarf radius $R_{\rm WD}=0.019\pm0.001~{\rm R_\odot}$, companion radius $R_B=0.118\pm0.006~{\rm R_\odot}$, and orbital inclination $i=80\pm1^\circ$, which corresponds to a companion mass $M_2=0.096\pm0.006~{\rm M_\odot}$. We display the companion mass-radius plot in Figure \ref{fig:2212_parsons2017_fig6} with the \citet{baraffe2015} low-mass stellar models over-plotted for various companion ages. These models suggest that the true inclination is $i_0\approx80^\circ$, in good agreement with the inclination estimates from our eclipsing light curve fits. We present our fitted parameter distributions in Figure \ref{fig:2212_apo_bg40_fit}.

We fit the PanSTARRS SED with a composite model SED to estimate the companion mass and confirm our previous mass estimate. Our best-fitting composite model returns a most probable companion mass $M_2,{\rm SED}=0.087^{+0.004}_{-0.002}~{\rm M_\odot}$. This SED mass estimate agrees within $1\sigma$ with our estimates obtained through radial velocity and eclipsing light curve fitting.

\section{Summary, Conclusions, and Future Prospects}
\input{tables/big_table.tex}

We identified and followed-up four new deeply-eclipsing, short period, white dwarf binaries with low-mass stellar companions using a box least squares periodicity search of objects in the Gaia DR2 white dwarf catalogue of \citet{gentilefusillo2019} with data from ZTF DR4. We fit our follow-up spectroscopy and determined the atmospheric parameters for each white dwarf. We obtained high-speed photometry of each system and place constraints on the companion mass and radius using the low-mass stellar models of \citet{baraffe2015}. Finally, we used time-series spectroscopy, together with our eclipsing light curve constraints, to obtain precise mass estimates for two of our binaries, J1844$+$4857 and J2212$+$5347. We fitted the available SED for each binary to place constraints on the masses of the companions of the remaining two binaries, J1644$+$2434 and J1744$+$3902, and confirm the companion masses for J1844$+$4857 and J2212$+$5347 obtained through radial velocity observations.  We present a summary of our numerical results in Table \ref{tab:big_table}.

While we do obtain mass and radius estimates for each of our binaries, we note that our estimates are dependent on low-mass stellar models. \citet{parsons2018} have shown that low-mass stellar models in the mass ranges of our objects under-estimate the true radius values by a few percent. Thus, our reported radii are potentially underestimated by a few percent. Deeper eclipsing light curves with better phase sampling are desired to obtain model-independent estimates for the masses and radii of the companion stars in these binaries. Additionally, phase-resolved spectroscopy surrounding the $H\alpha$ line for each of these binaries may show a reflected emission component within the core of the $H\alpha$ absorption line which would provide an independent method to confirm the mass ratio of the binary through the ratio of velocity semi-amplitudes \citep[see][]{parsons2017}.

\input{tables/wdm_candidates_table.tex}

Finally, we performed a preliminary search for similar systems within ZTF DR7 using the updated Gaia eDR3 white dwarf catalogue of \citet{gentilefusillo2021} as a base with the same BLS period-finding algorithm described above. Our preliminary search yields an additional 41 similar deeply-eclipsing light curves with periods ranging from $P=1.4-15.9~{\rm h}$, including 12 previously studied WD+M binaries, seven of which are known to be eclipsing.
We summarize these 41 additional eclipsing WD+M candidates in Table \ref{tab:wdm_candidates_table}. The continued study of these eclipsing WD+M binaries will assist in expanding the work of \citet{parsons2018} towards identifying the cause and magnitude of over-inflation in low-mass stellar objects and its potential relation to post-common-envelope evolution.

\subsection{Missed Eclipsing WD$+$BD Identifications}
For the four previously studied eclipsing white dwarf $+$ brown dwarf (WD$+$BD) binaries, our period-finding algorithm identifies ZTFJ0038+2030 \citep{vanroestel2021b} in the ZTF DR7 data archive, but assigns an incorrect period due to having only four in-eclipse data points. Our algorithm rejects the light curves of CSS21055 \citep{beuermann2013} and WD1032+011AB \citep{casewell2020} due to the light curves having too few $4\sigma$ or $5\sigma$ deviant points and fails to identify SDSS J1205$-$0242 \citep{parsons2017} due to its lack of public ZTF observations.

That our algorithm failed to identify two of the three known eclipsing WD$+$BD binaries with public data in ZTF suggests that other similar systems exist within the ZTF data archive of the \citet{gentilefusillo2021} sample of Gaia eDR3 white dwarfs. A more sophisticated algorithm with relaxed target cuts is thus required to identify these systems within the ZTF archive with more completeness. 

\input{figures/wd_bd_cmd.tex}
\input{figures/wd_bd_lc.tex}
\subsection{Two New Eclipsing White Dwarf $+$ Brown Dwarf Candidates}
To identify any potential new WD+BD binaries within our WD+M candidate list, we created a color-magnitude diagram using the PanSTARRS ($g-y$) color and absolute $g$-band magnitude, through the Gaia eDR3 parallax, for our 41 WD$+$M candidates. We present our color-magnitude diagram in Figure \ref{fig:wd_bd_cmd}. Open/closed black circles represent the 41 candidate/confirmed WD+M binaries presented in Table \ref{tab:wdm_candidates_table}. The four new WD$+$M systems presented in this work are marked as filled orange circles. We over-plot the 14 confirmed WD$+$BD binaries with blue star symbols. It is apparent that the WD$+$BD binaries form a cluster of points at fainter magnitudes and bluer colors due to the lack of near-IR contribution from their sub-stellar companions.

Among our 29 candidate WD$+$M binaries, two objects stand out as likely eclipsing WD$+$BD binaries: ZTF J110045.1473$+$521043.6242 (J1100$+$5210) and ZTF J182848.7659$+$230838.0508 (J1828$+$2308). We present the ZTF DR7 light curves for both of these WD$+$BD candidates in Figure \ref{fig:wd_bd_lc}, phase-folded to the most-probable period determined through our BLS period-finding algorithm. While the true eclipse depth of J1828$+$2308 is unmeasured in ZTF, the light curve of J1100$+$5210 shows a weak reflection effect and an eclipse shape that suggests a grazing-eclipse system geometry. Follow-up high-speed photometry and time-series spectroscopy observations are required to obtain estimates to the radii and masses for these systems. If confirmed, such observations would bring the number of eclipsing WD$+$BD binaries known to six from four, a small but significant increase.

\section*{Acknowledgements}

This work was supported in part by the NSF under grants AST-1906379 and AST-2107982, the Smithsonian Institution, the NSERC Canada, NASA under grants 80NSSC22K0338 and 80NSSC22K0479, and by the Fund FRQ-NT (Qu\'ebec).

TK acknowledges support from the National Science Foundation through grant AST 2107982, from NASA through grant 80NSSC22K0338, and from STScI through grant HST-GO-16659.002-A.

Based on observations obtained at the MMT, Apache Point Observatory, and the ZTF. The MMT is a joint facility of the Smithsonian Institution and the University of Arizona. The Apache Point Observatory 3.5-meter telescope is owned and operated by the Astrophysical Research Consortium. 

ZTF is supported by the National Science Foundation under Grant No. AST-1440341 and a collaboration including Caltech, IPAC, the Weizmann Institute for Science, the Oskar Klein Center at Stockholm University, the University of Maryland, the University of Washington, Deutsches Elektronen- Synchrotron and Humboldt University, Los Alamos National Laboratories, the TANGO Consortium of Taiwan, the University of Wisconsin at Milwaukee, and Lawrence Berkeley National Laboratories. Operations are conducted by COO, IPAC, and UW.

This research made use of Astropy\footnote{http://www.astropy.org}, a community-developed core Python package for Astronomy.

\section{Data Availability}
The data underlying this article will be shared on reasonable request to the corresponding author.




\input{references}



\bsp	
\label{lastpage}
\end{document}

%% file: figures/4_ztf_lightcurves.tex
  \begin{figure*}
    \hspace*{-1.35cm}
    \includegraphics[scale=0.675]{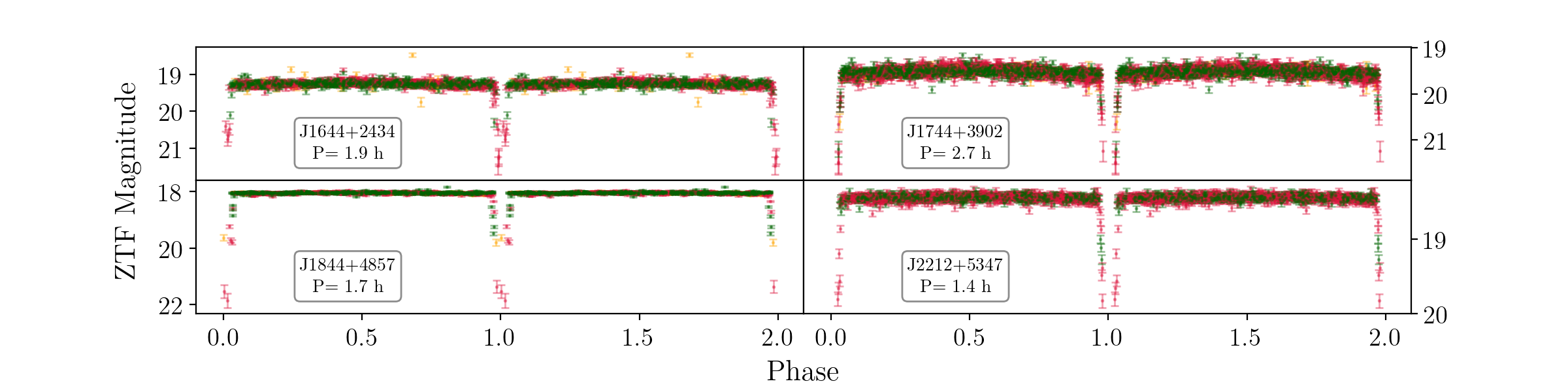}
    \caption{ZTF public DR4 light curves for the four deeply-eclipsing white dwarf binaries followed-up in this work. Each light curve has been phase-folded to its most probable period, obtained through a box least squares period-finding algorithm \citet{kovacs2002}. Individual data points are colored based on which filter they were measured in: green points used the ZTF $g$-band, red points used the ZTF $r$-band, and orange points used the ZTF $i$-band. Data across all filters have been median combined to the median value of the ZTF $g$-band filter.}
    \label{fig:4_ztf_lightcurves}
  \end{figure*}

%% file: figures/4_spectra.tex
\begin{figure*}
    \centering
    \includegraphics[scale=0.275]{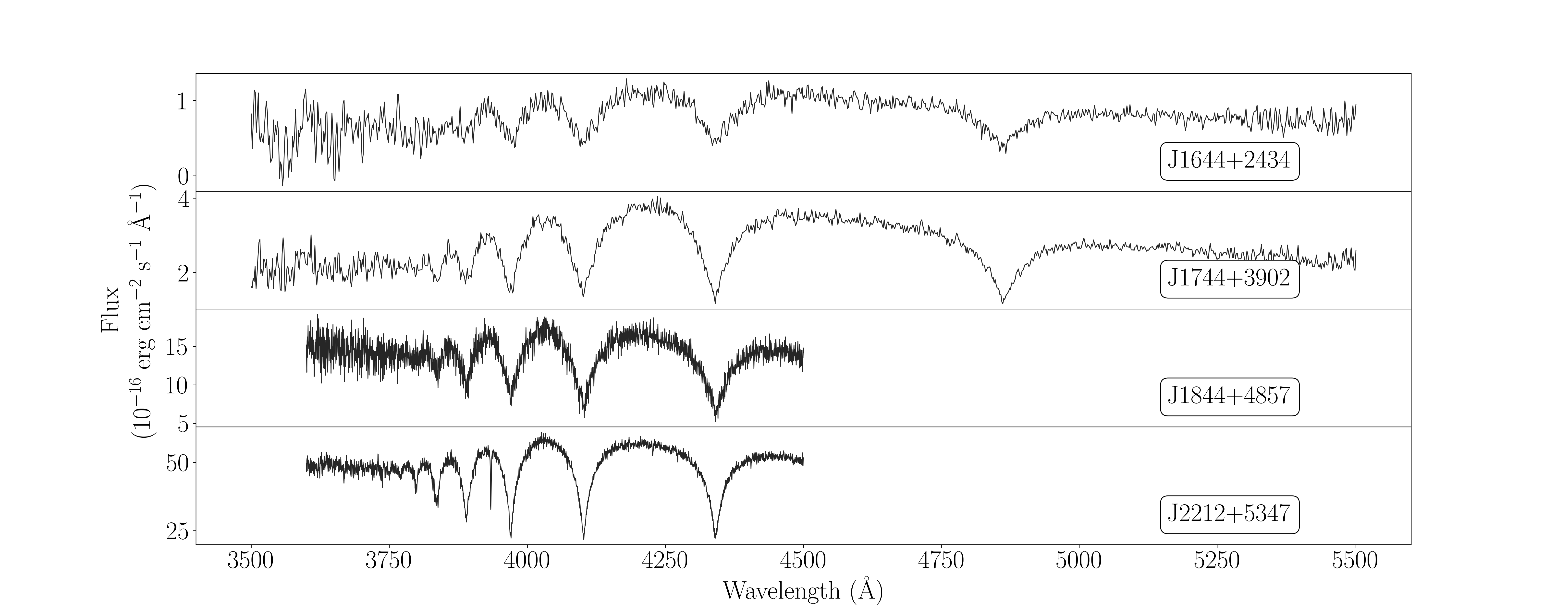}
    \caption{The APO 3.5m (top two panels) and the 6.5m MMT (bottom two panels) optical spectra for the four deeply eclipsing WD binaries identified in this work.}
    \label{fig:4_spectra}
\end{figure*}

%% file: figures/4_spectra_fits.tex
\begin{figure}
    \centering
    \includegraphics[scale=0.275]{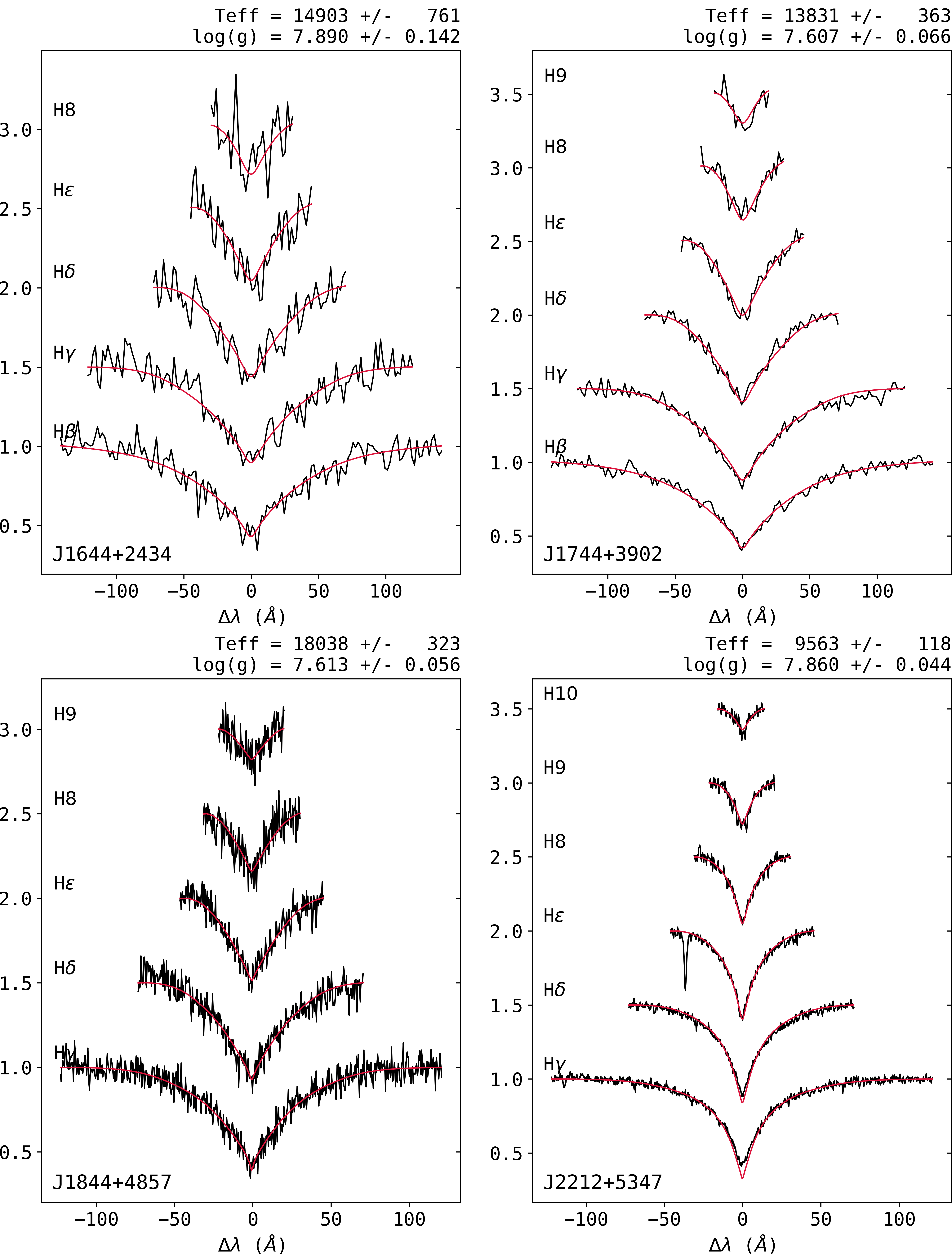}
    \caption{Pure-hydrogen single white dwarf model atmosphere fits to the four deeply-eclipsing white dwarfs presented in this work. The region surrounding the Calcium absorption feature at $\lambda\approx3933~{\rm Angstrom}$ is masked from our fits for J2212$+$5347. Best-fit uncertainties include the external uncertainties of $\sigma_{T_{\rm eff}}\approx1.2\%$ and $\sigma_{\log{g}}\approx0.038~{\rm dex}$ from \citet{liebert2005} added in quadrature. Best-fit parameters shown here have not been 3D corrected.}
    \label{fig:4_spectra_fits}
\end{figure}

%% file: figures/1644_apo_bg40_fit.tex
\begin{figure}
  \hspace*{0cm}
  \includegraphics[scale=0.25]{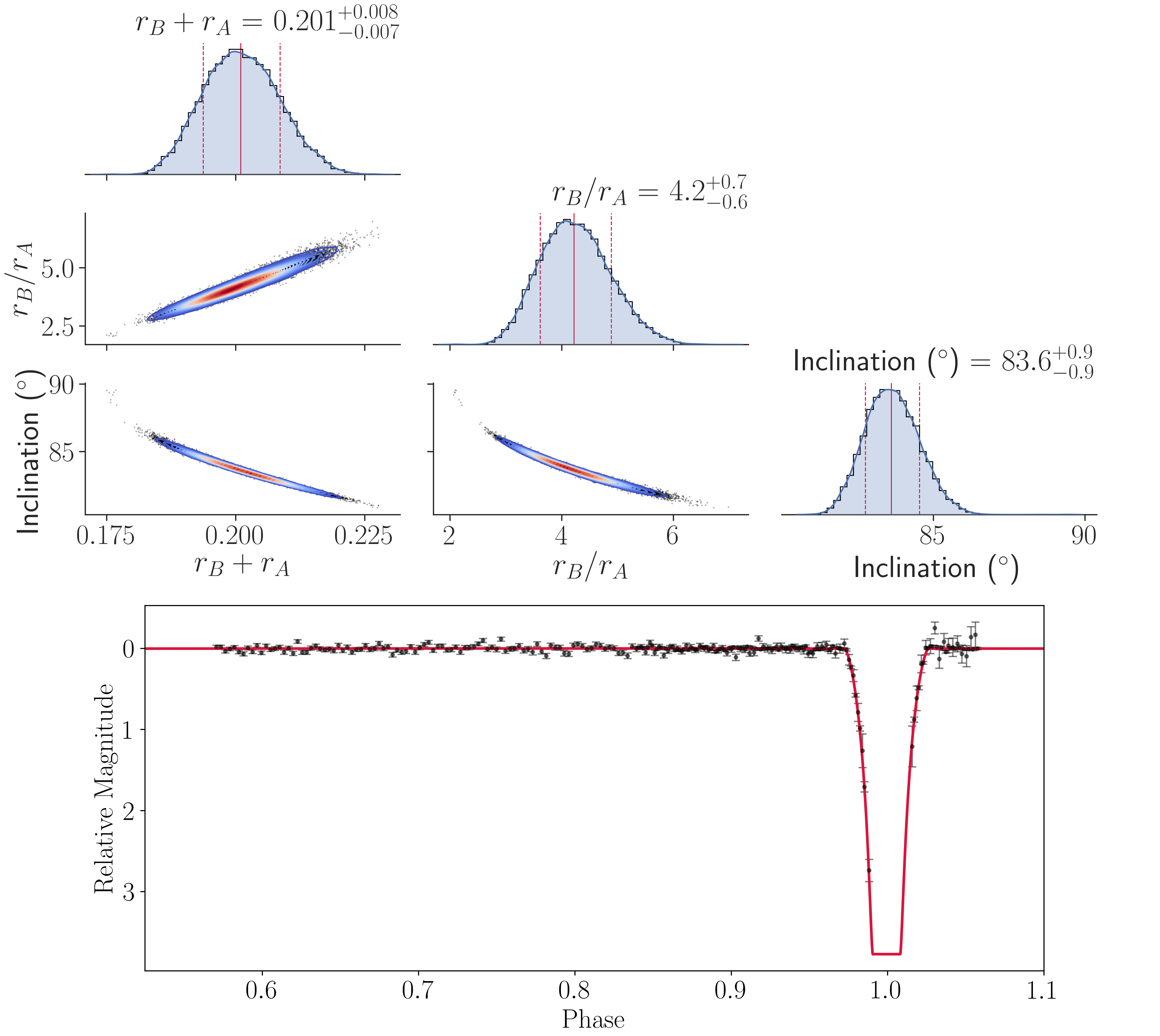}
  \caption{Top: Corner plot showing the 1D parameter distributions for our 10,000 Monte Carlo fits to the APO BG40 light curve of J1644$+$2434 on the diagonal and the 2D distributions on the off-diagonal. Bottom: APO BG40 light curve for J1644$+$2434 (black) with the best-fit model over-plotted in red.}
  \label{fig:1644_apo_bg40_fit}
\end{figure}

%% file: figures/2_sed_models.tex
\begin{figure}
  \includegraphics[scale=0.4]{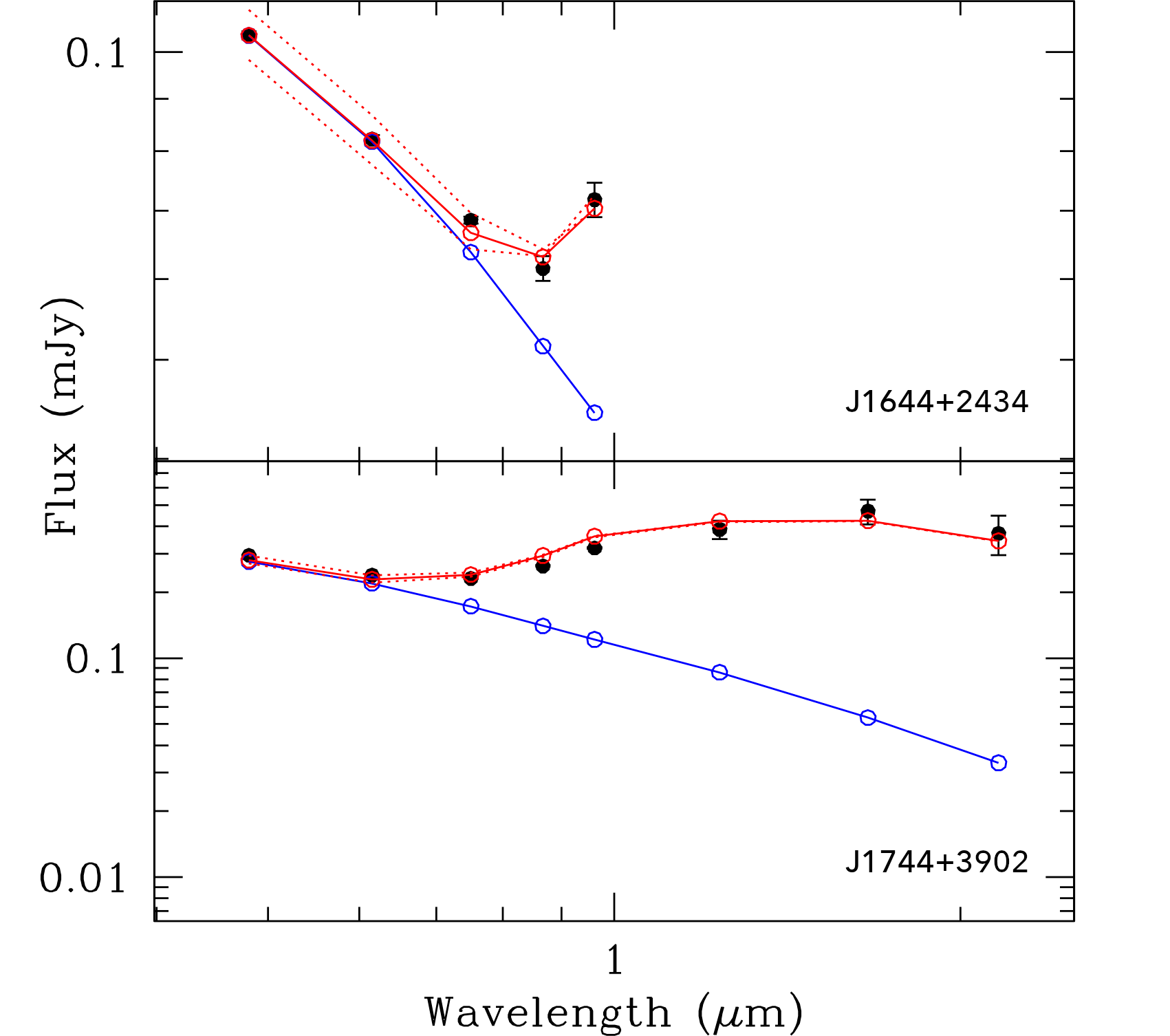}
  \caption{Spectral energy distributions for two deeply-eclipsing white dwarf binaries followed-up in this work, J1644$+$2434 (top) and J1744$+$3902 (bottom). Black points represent measured fluxes in various bands. The blue curve represents the SED of a single DA white dwarf using parameters derived through our spectroscopic fitting. The red solid and dotted curves represents the best-fitting composite SED and its uncertainty based on Gaia eDR3 distances and \citet{baraffe2015} low-mass stellar model colors.}
  \label{fig:2_sed_models}
\end{figure}

%% file: figures/1744_apo_rband_fit2.tex
\begin{figure}
  \hspace*{0cm}
  \includegraphics[scale=0.25]{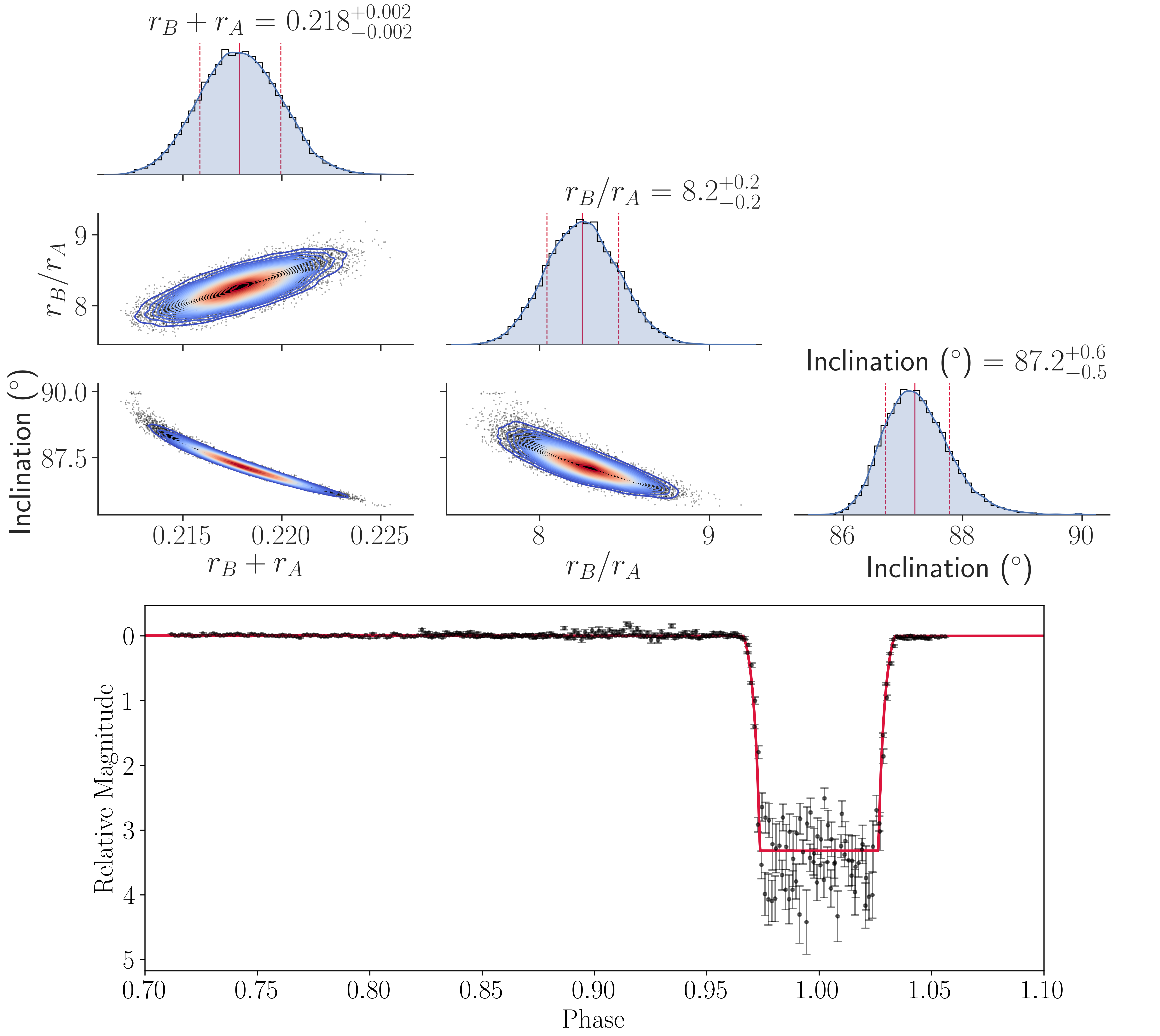}
  \caption{Top: Corner plot showing the 1-D parameter distributions for our 10,000 Monte Carlo fits to the APO $r$-band light curve of J1744$+$3902 on the diagonal and the 2-D distributions on the off-diagonal. Bottom: APO $r$-band light curve for J1744$+$3902 (black) with the best-fit model over-plotted in red.}
  \label{fig:1744_apo_rband_fit2}
\end{figure}

%% file: figures/1744_apo_iband_fit2.tex
\begin{figure}
  \hspace*{0cm}
  \includegraphics[scale=0.25]{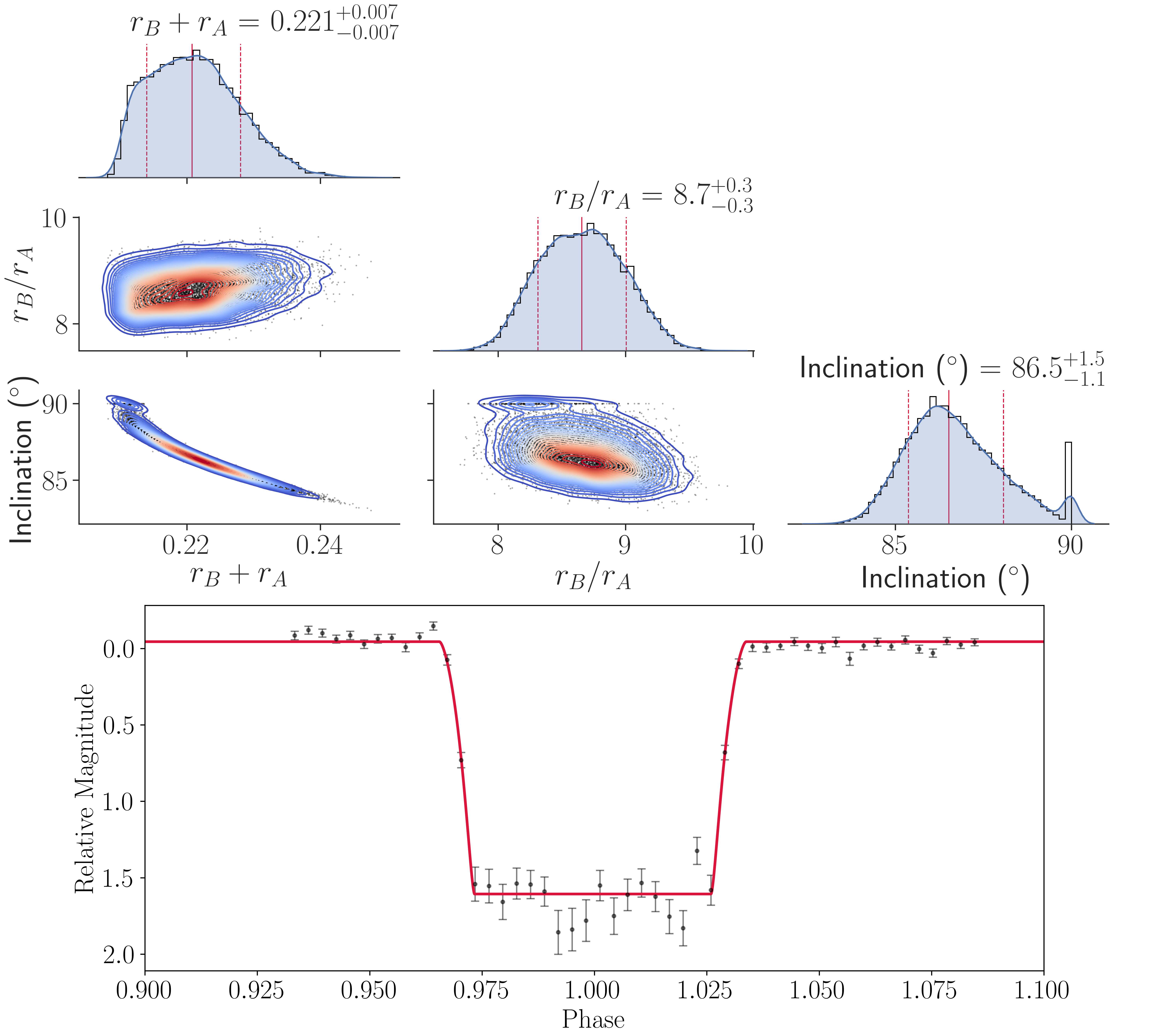}
  \caption{Top: Corner plot showing the 1-D parameter distributions for our 10,000 Monte Carlo fits to the APO $i$-band light curve of J1744$+$3902 on the diagonal and the 2-D distributions on the off-diagonal. Bottom: APO $i$-band light curve for J1744$+$3902 (black) with the best-fit model over-plotted in red.}
  \label{fig:1744_apo_iband_fit2}
\end{figure}

%% file: figures/1844_mmt_rv_figure.tex
\begin{figure}
  \includegraphics[scale=0.4]{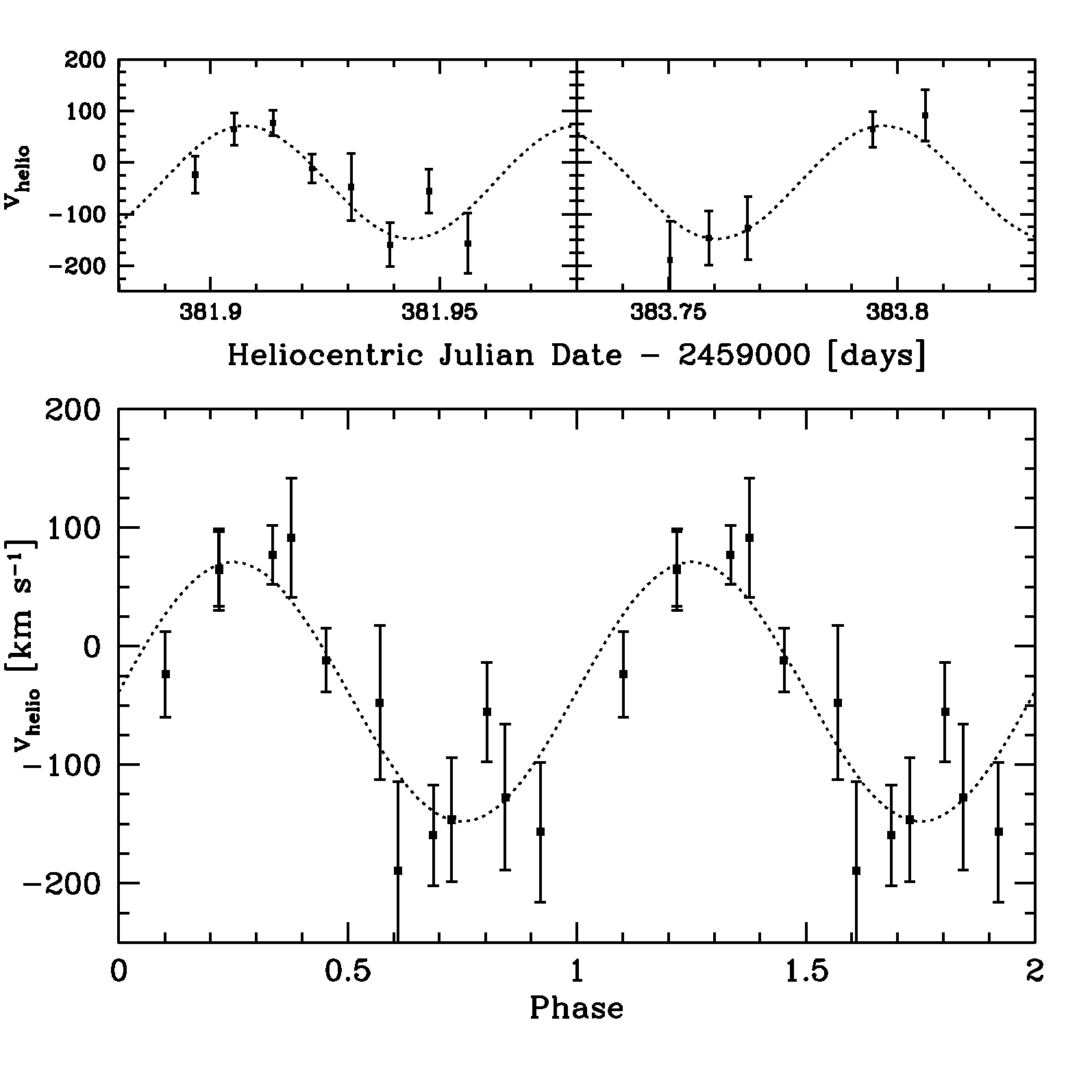}
  \caption{Best-fitting circular orbit to J1844$+$4857. We fix the period at the photometric value obtained from a baseline of over 800 days of ZTF observations. Individual radial velocity measurements are presented in Table \ref{tab:rv_table_1844_2212}}
  \label{fig:1844_mmt_rv_figure}
\end{figure}

%% file: figures/1844_parsons2017_fig6.tex
\begin{figure}
  \hspace{0.2in}
  \includegraphics[scale=0.45]{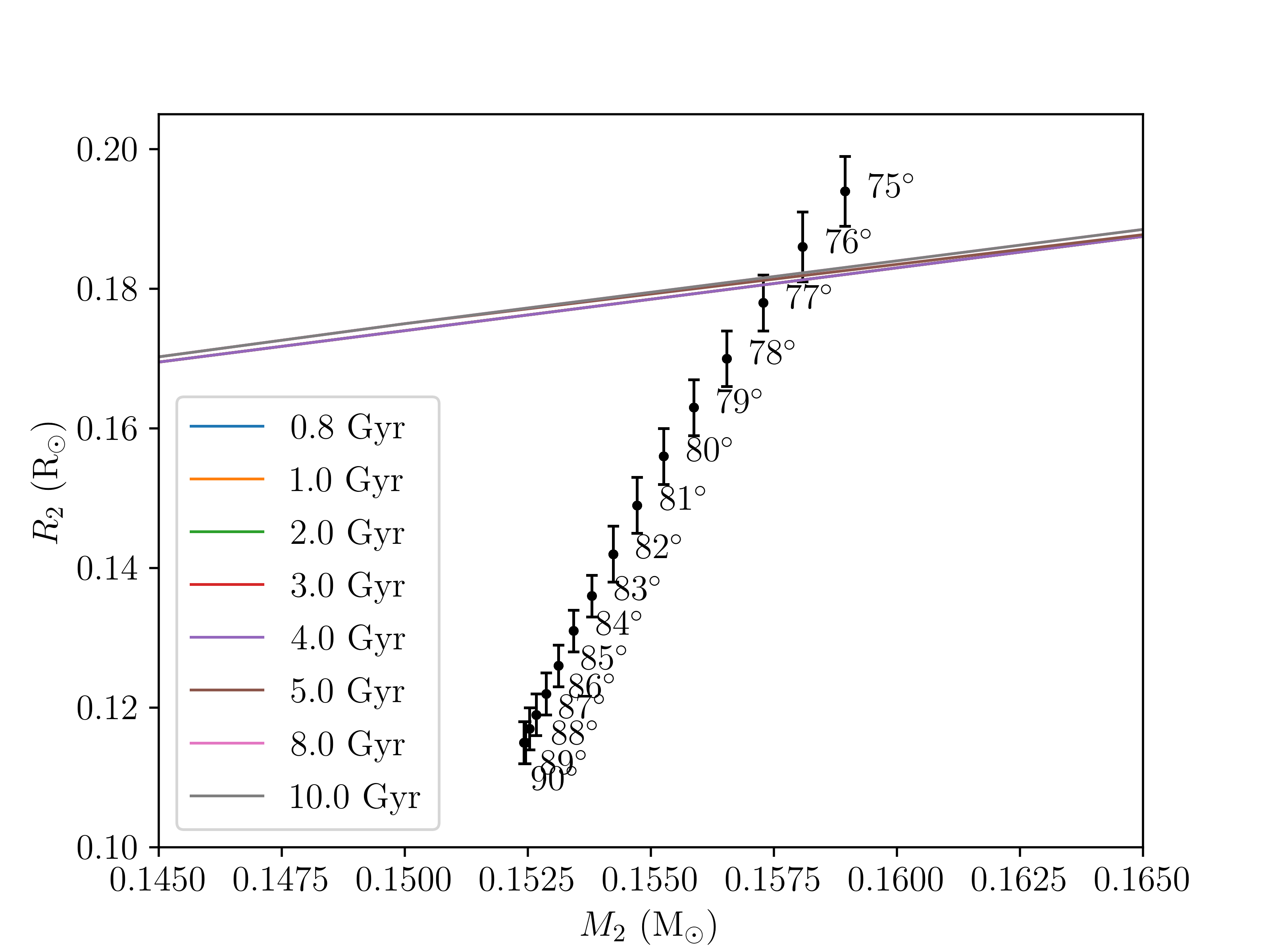}
  \caption{Companion radius vs. mass plot for J1844$+$4857. The black data points mark the best-fitting JKTEBOP model light curve parameters to the APO BG40 light curve of J1844$+$4857 for a fixed inclination. The low-mass stellar models of \citet{baraffe2015} for varying system ages are overplotted as colored lines.}
  \label{fig:1844_parsons2017_fig6}
\end{figure}

%% file: figures/1844_apo_bg40_fit.tex
\begin{figure}
  \hspace*{0cm}
  \includegraphics[scale=0.225]{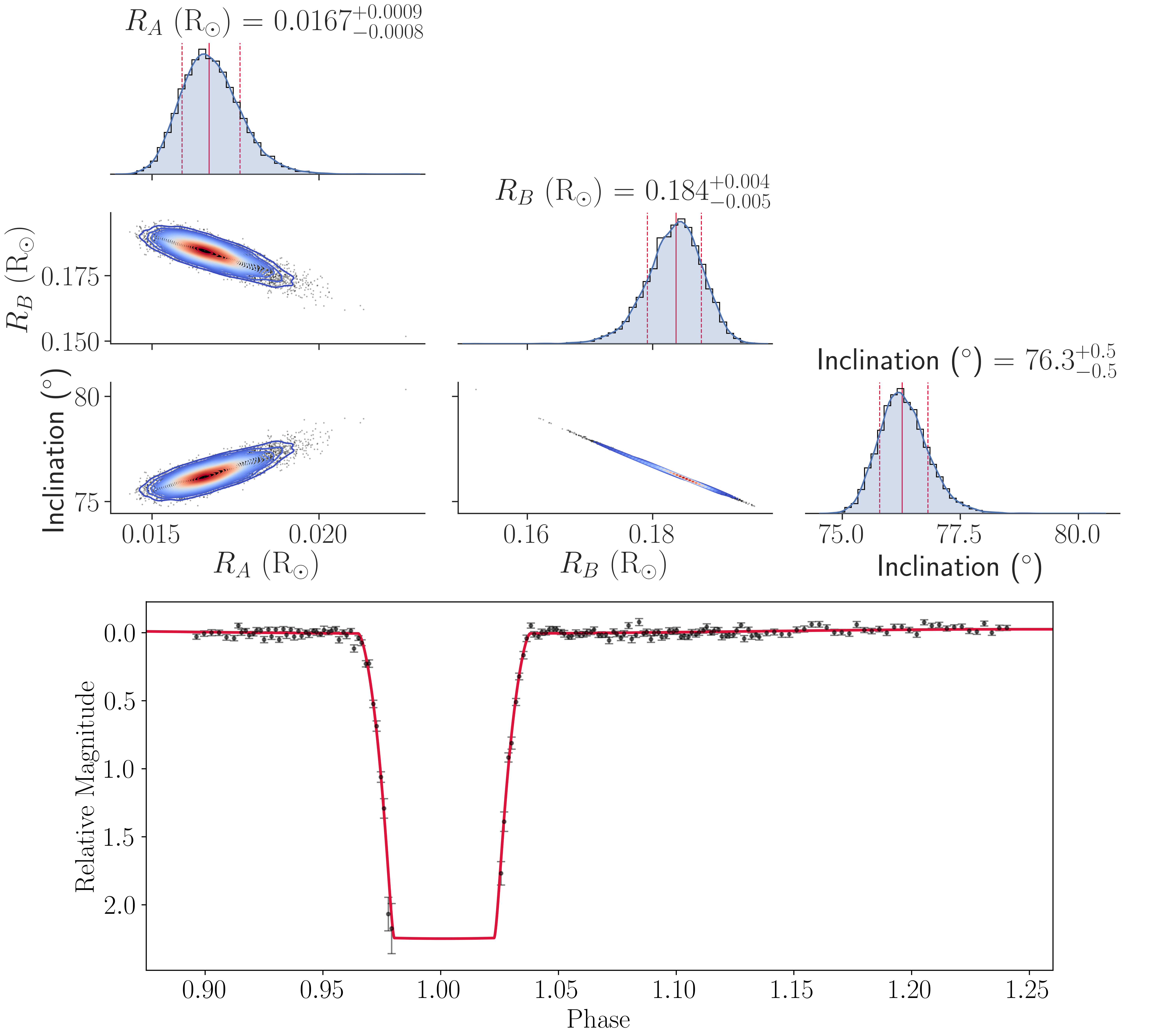}
  \caption{Top: Corner plot showing the 1D parameter distributions for our 10,000 Monte Carlo fits to the APO BG40 light curve of J1844$+$4857 on the diagonal and the 2D distributions on the off-diagonal. Bottom: APO BG40 light curve for J1844$+$4857 (black) with the best-fit model overplotted in red.}
  \label{fig:1844_apo_bg40_fit}
\end{figure}

%% file: tables/rv_table_1844_2212.tex
\begin{table}
    \centering
    \begin{tabular}{c|c|c|c}
  \hline
  \hline
  {Object Name} & {HJD} & {Radial Velocity} & {Error} \\
  {} & {($-2450000~{\rm d}$)} & (${\rm km~s^{-1}}$) & (${\rm km~s^{-1}}$) \\
  \hline
  \hline
  {J1844$+$4857} & {9381.896673} & { -23.74} & {36.18} \\
  {} & {9381.905175} & {  65.11} & {31.55} \\
  {} & {9381.913675} & {  77.01} & {24.82} \\
  {} & {9381.922178} & { -11.66} & {27.05} \\
  {} & {9381.930686} & { -47.46} & {64.94} \\
  {} & {9381.939184} & {-159.47} & {42.53} \\ 
  {} & {9381.947686} & { -55.44} & {42.00} \\
  {} & {9381.956192} & {-156.77} & {58.82} \\
  {} & {9383.750346} & {-189.51} & {75.27} \\
  {} & {9383.758844} & {-146.40} & {52.36} \\
  {} & {9383.767346} & {-127.26} & {61.61} \\
  {} & {9383.794594} & {  64.45} & {34.45} \\
  {} & {9383.806067} & {  91.44} & {50.24} \\
  \hline
  {J2212$+$5347} & {9192.641711} & {-75.47} & {13.98} \\
  {} & {9192.646743} & {-114.66} & {16.15} \\
  {} & {9192.651777} & {-99.90} & {13.68} \\
  {} & {9192.656807} & {-105.96} & {14.24} \\
  {} & {9192.661835} & {-52.60} & {17.92} \\
  {} & {9192.666860} & {-8.11} & {16.05} \\
  {} & {9192.671886} & {-2.83} & {18.13} \\
  {} & {9192.676913} & {24.14} & {17.06} \\
  {} & {9192.682285} & {16.91} & {19.68} \\
  {} & {9192.688723} & {-8.18} & {16.37} \\
  {} & {9192.696543} & {-44.29} & {23.46} \\
  {} & {9195.652166} & {46.51} & {12.50} \\
  {} & {9195.659970} & {48.98} & {14.81} \\
  {} & {9195.668124} & {67.02} & {50.40} \\
  {} & {9198.644382} & {-30.09} & {29.73} \\
  {} & {9198.652949} & {-76.97} & {25.88} \\
  {} & {9199.597496} & {-113.44} & {14.12} \\
  {} & {9199.603251} & {-70.99} & {13.93} \\
  {} & {9199.608980} & {4.67} & {15.24} \\
  {} & {9199.614717} & {39.12} & {14.62} \\
  {} & {9199.620450} & {60.72} & {12.85} \\
  {} & {9199.626181} & {20.96} & {17.38} \\
  {} & {9199.631917} & {-0.69} & {12.54} \\
  {} & {9199.637655} & {-34.63} & {19.78} \\
  {} & {9199.643389} & {-100.21} & {14.21} \\
  {} & {9199.649134} & {-120.65} & {13.23} \\
  {} & {9199.654876} & {-104.55} & {14.16} \\
  \hline
  \hline
    \end{tabular}
    \caption{Radial velocity measurements of J1844$+$4857 and J2212$+$5347.}

    \label{tab:rv_table_1844_2212}
\end{table}

%% file: figures/2212_mmt_rv_figure.tex
\begin{figure}
  \includegraphics[scale=0.4]{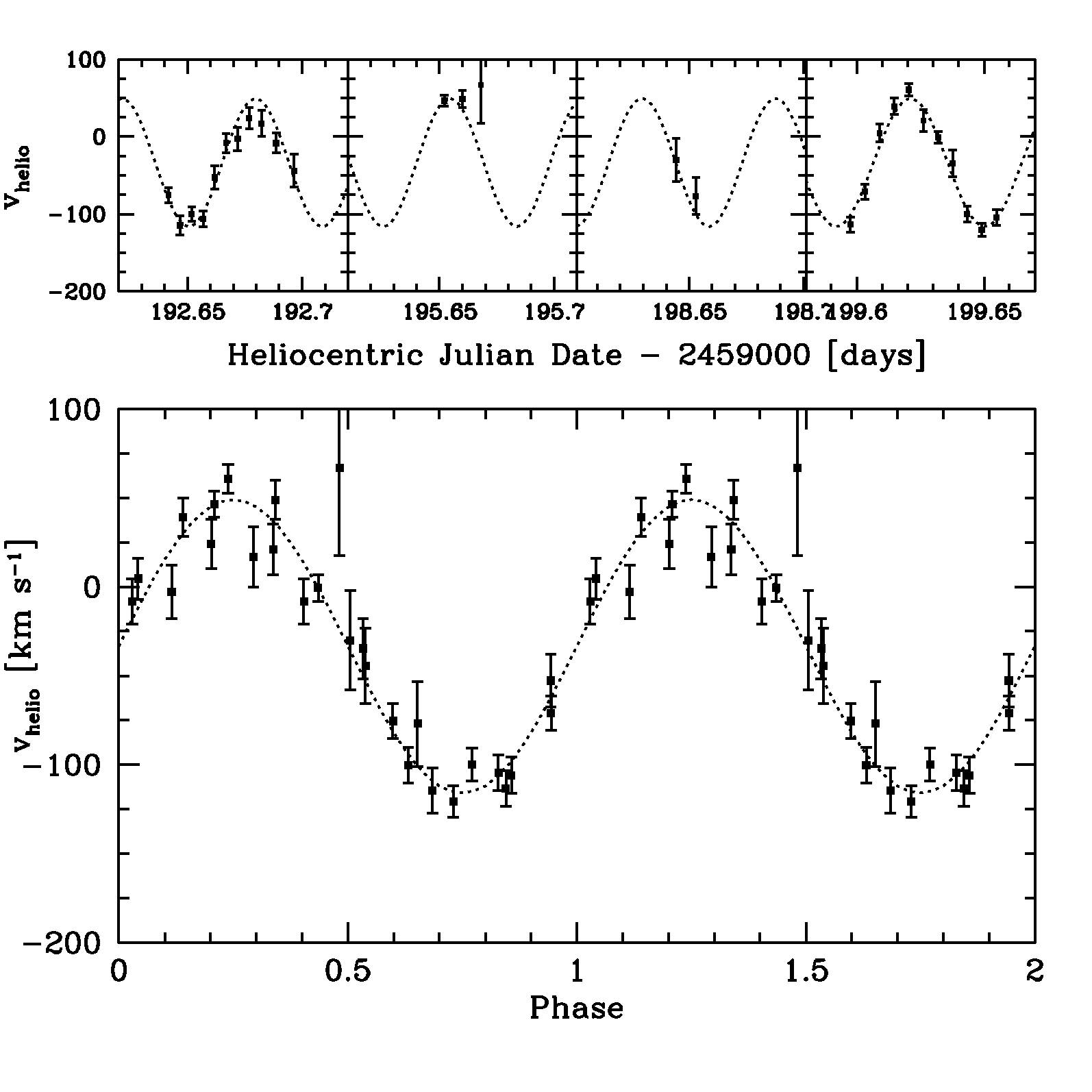}
  \caption{Best-fitting circular orbit to J2212$+$5347. Individual radial velocity measurements are presented in Table \ref{tab:rv_table_1844_2212}}
  \label{fig:2212_mmt_rv_figure}
\end{figure}

%% file: figures/2212_parsons2017_fig6.tex
\begin{figure}
  \hspace{0.2in}
  \includegraphics[scale=0.45]{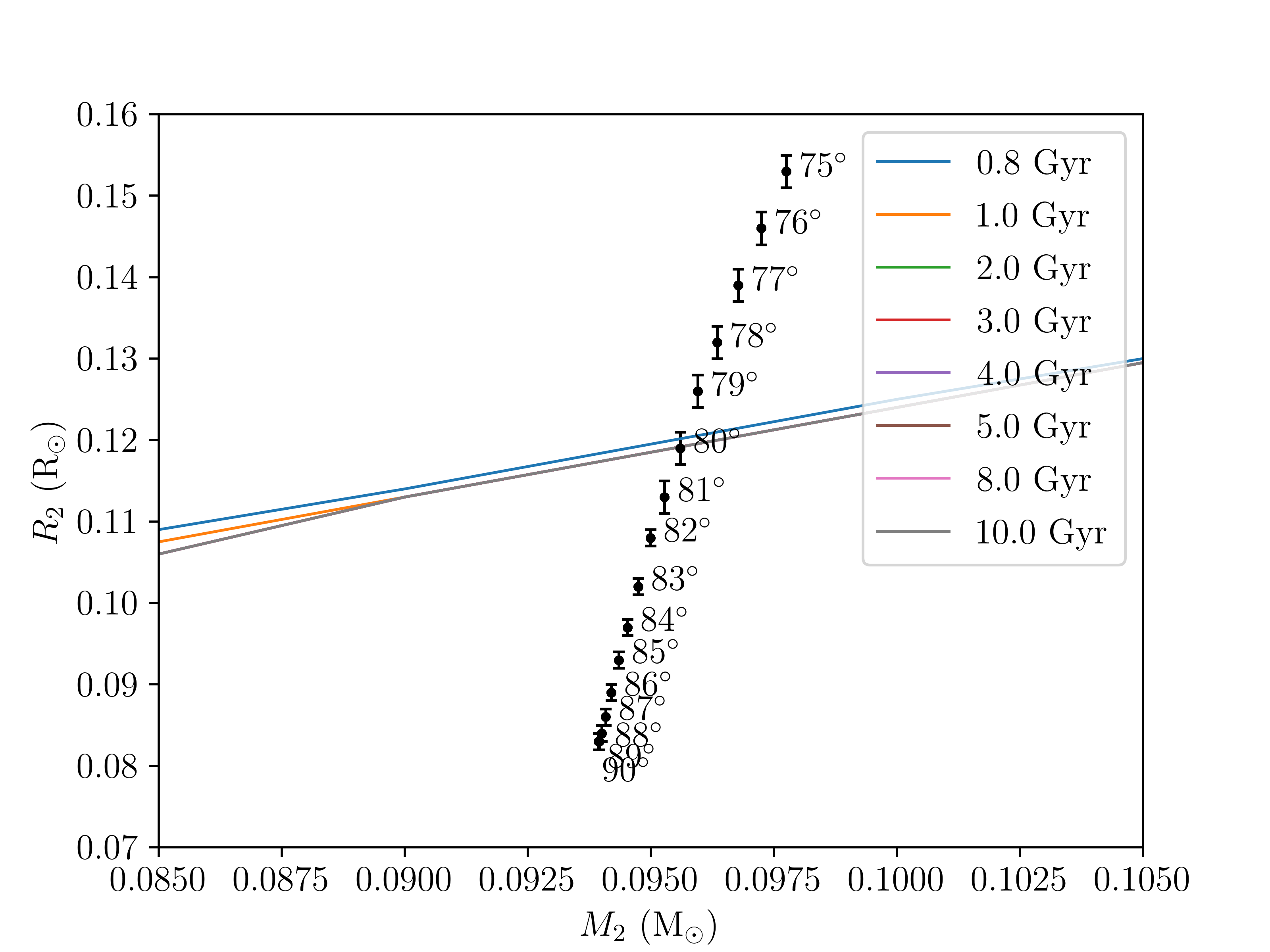}
  \caption{Companion radius vs. mass plot for J2212$+$5347. The black data points mark the best-fitting JKTEBOP model light curve parameters to the APO BG40 light curve of J2212$+$5347 for a fixed inclination. The low-mass stellar models of \citet{baraffe2015} for varying system ages are overplotted as colored lines.}
  \label{fig:2212_parsons2017_fig6}
\end{figure}

%% file: figures/2212_apo_bg40_fit.tex
\begin{figure}
  \includegraphics[scale=0.25]{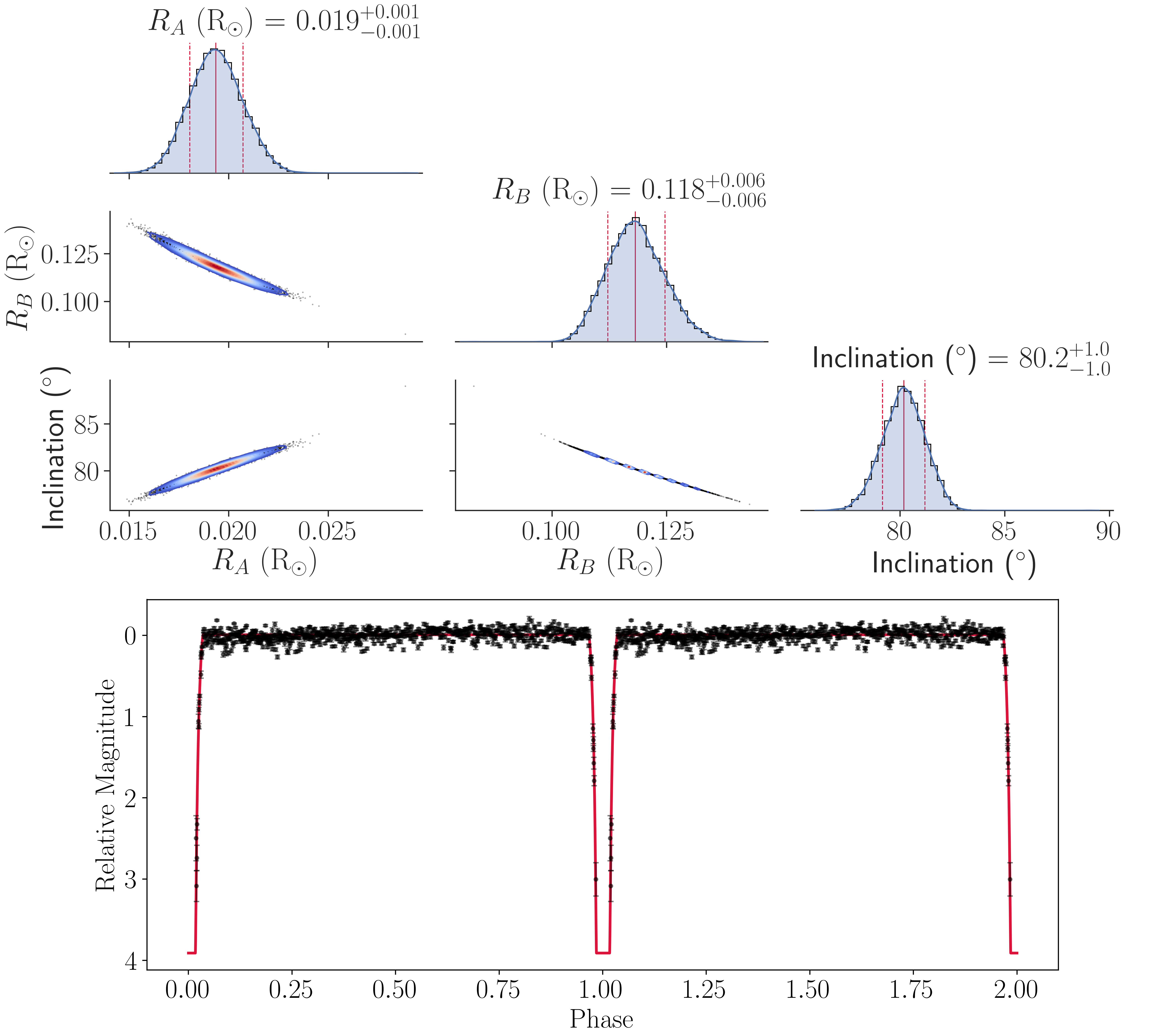}
  \caption{Top: Corner plot presenting the distribution for each fitted parameter. The diagonal entries display the 1D distributions while the off-diagonal entries display the 2D correlation plots between each fitted parameter. Median parameter values with 15.87 and 84.13 percentile uncertainties are presented above each 1D distribution. Bottom: APO BG40 light curve for J2212$+$5347 (black) with the best-fit model over-plotted in red.}
  \label{fig:2212_apo_bg40_fit}
\end{figure}

%% file: tables/big_table.tex
\begin{table*}
	\centering
  \small
  \renewcommand{\arraystretch}{1.5}
  \addtolength{\tabcolsep}{2pt}
	\begin{tabular}{l c c c c}
    \hline
    \hline
    {} & J1644+2434 & J1744$+$3902 & J1844$+$4857 & J2212+5347 \\
    \hline
    Gaia ID (eDR3) & 1300187622427241856 & 1343069434903597952 & 2119978952315202176 & 2004624931143291648 \\
    R.A. (J2000) & 16:44:41.1946 & 17:44:24.7141 & 18:44:34.3978 & 22:12:26.9672 \\
    Decl. (J2000) & +24:34:28.2112 & +39:02:15.6653 & +48:57:36.5063 & +53:47:50.6967 \\
    Gaia G (mag) & 19.1 & 17.8 & 19.3 & 18.3 \\
    Gaia Parallax (mas) & $2.43\pm0.22$ & $4.14\pm0.09$ & $1.26\pm0.21$ & $5.12\pm0.11$ \\
    Period (min) & 115.35 & 161.97 & 104.64 & 84.01 \\
    \hline
    $T_{\rm eff}$ (${\rm K}$) & $14900\pm760$ & $13830\pm360$ & $18040\pm320$ & $9430\pm120$ \\
    $\log{g}$ (cgs) & $7.89\pm0.14$ & $7.61\pm0.07$ & $7.61\pm0.06$ & $7.56\pm0.04$ \\
    $M_{\rm WD,CO}$ (${\rm M_\odot}$) & $0.55\pm0.07$ & $0.41\pm0.03$ & $0.43\pm0.02$ & $0.38\pm0.02$ \\
    $R_{\rm WD}$ (${\rm R_\odot}$) & $0.014\pm0.002$ & $0.017\pm0.001$ & $0.017\pm0.001$ & $0.017\pm0.001$ \\
    \hline
    $\frac{R_B+R_A}{a}$ & $>0.201^{+0.008}_{-0.007}$ & $0.218\pm0.002$ & $0.324\pm0.006$ & $0.27\pm0.01$ \\
    $\frac{R_B}{R_A}$ & $>4.2^{+0.7}_{-0.6}$ & $8.2\pm0.2$ & $11.0\pm0.8$ &$6.1^{+0.8}_{-0.7}$  \\
    Inclination ($^\circ$) & $<83.6\pm0.9$ & $87.2^{+0.6}_{-0.5}$ & $76.3\pm0.5$ & $80\pm1$ \\
    $a$ (${\rm R_\odot}$) & $>0.36\pm0.07$ & $0.72\pm0.04$ & $0.62\pm0.02^\dagger$ & $0.502\pm0.007^\dagger$ \\
    $R_{\rm A}$ (${\rm R_\odot}$) & . . . & . . . & $0.0167^{+0.0009}_{-0.0008}$ & $0.019\pm0.001$ \\
    $R_B$ (${\rm R_\odot}$) & $>0.06\pm0.01$ & $0.14\pm0.01$ & $0.184^{+0.004}_{-0.005}$ & $0.118\pm0.006$ \\
    $M_{B,{\rm RV}}$ (${\rm M_\odot}$) & . . . & . . . & $0.16\pm0.04$ & $0.096\pm0.006$ \\
    \hline
    $M_{\rm 2,SED}$ (${\rm M_\odot}$) & $0.084^{+0.004}_{-0.003}$ & $0.121^{+0.006}_{-0.007}$ & $0.137^{+0.010}_{-0.009}$ & $0.087^{+0.004}_{-0.002}$ \\
    \hline
    \hline
	\end{tabular}
    \caption{System parameters to the four WD+M binaries identified in this work. External atmospheric parameter uncertainties of $\sigma_{T_{\rm eff}}\approx1.2\%$ and $\sigma_{\log{g}}\approx0.038~{\rm dex}$ from \citet{liebert2005} have been added in quadrature with the systematic uncertainties from our spectroscopic fits. Spectroscopic white dwarf masses are based on CO-core composition. Values marked with $^\dagger$ were calculated using combined results from spectroscopic and light curve fitting.}
    \label{tab:big_table}
\end{table*}

%% file: tables/wdm_candidates_table.tex
\begin{table*}
  \centering
  \begin{tabular}{l|c|c|c|c|c|c|c}
  \hline
  \hline
  {Gaia ID} & {R.A.} & {Decl.} & {Gaia G} & {Gaia Parallax} & {BP$-$RP} & {ZTF Period} & {Original Work} \\
  {(eDR3)} & {(J2000)} & {(J2000)} & {(mag)} & {(mas)} & {(mag)} & {(h)} & {} \\
  \hline
2857103195527412992 & 00:22:34.4830 & +28:20:43.6224 & $18.78$ & $ 1.73\pm0.22$ & $ 0.67$ & $ 7.53$ & . . .\\
380585951271107968 & 00:33:35.3569 & +39:17:06.5496 & $18.84$ & $ 1.90\pm0.24$ & $ 0.16$ & $ 5.99$ & . . .\\
380560941677424768 & 00:33:52.6285 & +38:55:29.6483 & $18.34$ & $ 1.19\pm0.17$ & $-0.19$ & $ 4.86$ & . . .\\
2858607774110350976 & 00:37:32.3001 & +30:35:18.4958 & $18.77$ & $ 2.09\pm0.19$ & $ 0.66$ & $ 8.56$ & . . .\\
364169452395186304 & 00:50:20.3997 & +36:22:49.2657 & $18.00$ & $ 3.36\pm0.13$ & $ 0.82$ & $ 6.66$ & . . .\\
2584756467429594880$^{EcSp}$ & 01:10:09.1576 & +13:26:15.9011 & $16.71$ & $ 3.66\pm0.07$ & $-0.03$ & $ 3.99$ & \citep{silvestri2006,pyrzas2009} \\
320209019011329408 & 01:18:24.5125 & +34:45:04.2501 & $19.59$ & $ 1.40\pm0.45$ & $-0.17$ & $ 1.97$ & . . .\\
353846172081048960 & 02:19:19.7793 & +46:23:44.9523 & $19.73$ & $ 1.58\pm0.39$ & $ 0.57$ & $ 1.40$ & . . .\\
114286056089620096 & 02:48:53.2578 & +24:50:32.3459 & $20.00$ & $ 2.15\pm0.49$ & $ 0.57$ & $ 3.30$ & . . .\\
173280902235344640 & 04:29:55.3185 & +34:47:34.0032 & $17.01$ & $ 4.24\pm0.10$ & $ 0.70$ & $ 7.37$ & . . .\\
502209190190665856 & 05:06:33.6633 & +73:06:46.6020 & $18.62$ & $ 1.34\pm0.18$ & $ 0.40$ & $ 3.40$ & . . .\\
3352706275042496896 & 06:42:42.3970 & +13:14:27.7701 & $18.31$ & $ 3.77\pm0.20$ & $ 0.86$ & $ 4.11$ & . . .\\
1035619453763003904$^{Sp}$ & 08:20:11.2288 & +58:42:35.0730 & $18.42$ & $ 3.77\pm0.14$ & $ 0.86$ & $ 8.18$ &  \citep{liu2012}\\
661256399005907456$^{EcSp}$ & 08:38:45.8803 & +19:14:15.9461 & $18.17$ & $ 2.97\pm0.17$ & $ 0.36$ & $ 3.12$ & \citep{heller2009,parsons2013}\\
580790014913812608$^{EcSp}$ & 09:08:12.0396 & +06:04:21.1104 & $17.07$ & $ 3.71\pm0.09$ & $ 0.57$ & $ 3.59$ & \citep{silvestri2006,parsons2013}\\
794184674743537152$^{EcSp}$ & 09:39:47.9218 & +32:58:07.2437 & $17.79$ & $ 2.37\pm0.12$ & $ 0.07$ & $ 7.94$ & \citep{silvestri2007,parsons2013}\\
836510989033113472$^\dagger$ & 11:00:45.1473 & +52:10:43.6242 & $18.53$ & $ 1.57\pm0.15$ & $-0.24$ & $ 1.61$ & . . .\\
1055977736185041024$^{EcSp}$ & 11:08:05.3914 & +65:22:11.4435 & $18.50$ & $ 1.93\pm0.16$ & $ 0.21$ & $ 7.70$ & \citep{silvestri2006,rowan2019}\\
1680466809952508544$^{Sp}$ & 12:39:03.0139 & +65:49:34.3481 & $16.80$ & $ 4.36\pm0.05$ & $ 0.74$ & $15.89$ & \citep{silvestri2006} \\
1453111749770736256$^{Ec}$ & 14:08:47.1691 & +29:50:44.8266 & $18.64$ & $ 1.85\pm0.18$ & $ 0.10$ & $ 4.60$ & \citep{parsons2013}\\
1668534527515116416$^{Sp}$ & 14:12:20.7214 & +65:41:23.3975 & $18.94$ & $ 2.48\pm0.16$ & $ 0.82$ & $ 3.80$ & \citep{raymond2003}\\
1605099028784731776$^{Sp}$ & 14:29:53.3992 & +52:23:54.0114 & $19.48$ & $ 2.11\pm0.23$ & $ 0.48$ & $ 1.76$ & \citep{rebassa-mansergas2016}\\
1486587588165368576 & 14:35:47.8471 & +37:33:38.1632 & $17.03$ & $ 5.44\pm0.06$ & $ 0.64$ & $ 3.02$ & . . .\\
1383821420340997248$^{EcSp}$ & 15:48:45.9939 & +40:57:28.3554 & $18.20$ & $ 4.66\pm0.10$ & $ 0.65$ & $ 4.45$ & \citep{silvestri2006,pyrzas2009}\\
1702389319463842560 & 16:02:34.2133 & +72:52:09.0725 & $17.60$ & $ 5.80\pm0.07$ & $ 1.02$ & $ 4.84$ & . . .\\
1414195708232098688 & 17:13:36.4362 & +49:07:38.7396 & $20.12$ & $ 1.96\pm0.41$ & $ 0.71$ & $ 1.65$ & . . .\\
1438919253678919040 & 17:24:00.0579 & +60:44:51.8733 & $20.03$ & $ 1.45\pm0.47$ & $ 0.30$ & $ 2.13$ & . . .\\
4490536776397581312 & 17:25:23.9745 & +08:46:39.2603 & $18.89$ & $ 3.82\pm0.23$ & $ 0.77$ & $10.70$ & . . .\\
1633921248638338688 & 17:31:28.7554 & +66:38:11.4812 & $19.94$ & $ 1.48\pm0.56$ & $ 0.80$ & $ 3.75$ & . . .\\
4529477698683435264$^\dagger$ & 18:28:48.7659 & +23:08:38.0508 & $18.00$ & $ 4.91\pm0.10$ & $-0.11$ & $ 2.69$ & . . .\\
4198059364790247680 & 19:16:01.4188 & -12:34:10.9754 & $19.22$ & $ 2.66\pm0.37$ & $ 0.81$ & $ 3.30$ & . . .\\
2025873096433233664 & 19:20:14.1420 & +27:22:18.0851 & $15.56$ & $ 5.22\pm0.03$ & $ 0.31$ & $ 3.58$ & . . .\\
2045226627079102848 & 19:36:07.5619 & +31:55:06.5214 & $17.96$ & $ 4.75\pm0.10$ & $ 0.51$ & $ 4.19$ & . . .\\
2138404091435577216 & 19:36:19.9494 & +54:09:20.6823 & $18.48$ & $ 3.87\pm0.11$ & $ 0.26$ & $ 3.78$ & . . .\\
2238908319023604352 & 19:42:17.5053 & +59:39:45.7374 & $19.82$ & $ 1.38\pm0.31$ & $ 0.81$ & $ 6.64$ & . . .\\
1836736545022514560 & 20:06:41.3767 & +27:28:50.7222 & $18.68$ & $ 2.01\pm0.16$ & $ 0.36$ & $ 3.99$ & . . .\\
1805534569689099392 & 20:29:29.2711 & +15:21:03.7944 & $18.11$ & $ 3.03\pm0.15$ & $ 0.70$ & $ 3.31$ & . . .\\
1859639715829026432 & 20:49:48.5527 & +31:11:52.1649 & $19.07$ & $ 0.96\pm0.24$ & $ 0.56$ & $ 2.21$ & . . .\\
1795836842772548224 & 21:50:37.0243 & +23:40:00.1203 & $19.11$ & $ 0.56\pm0.31$ & $ 0.17$ & $ 6.39$ & . . .\\
2014898802148215424 & 23:08:28.5607 & +61:25:37.3785 & $19.41$ & $ 1.53\pm0.25$ & $ 0.85$ & $ 5.60$ & . . .\\
1926813725890052352$^{Sp}$ & 23:41:43.8924 & +45:24:31.7063 & $18.03$ & $ 2.40\pm0.11$ & $-0.04$ & $ 6.98$ & \citep{rebassa-mansergas2010}\\
  \hline
  \hline
  \end{tabular}
  \caption{Collection of the additional 41 eclipsing WD+M candidates identified in a preliminary ZTF DR7 search using box least squares periodicity search on the \citet{gentilefusillo2021} Gaia eDR3 white dwarf catalogue. Previously studied systems are labelled with $^{Ec}$ and/or $^{Sp}$ if eclipsing light curve or spectroscopic observations were obtained in their original works. The two targets marked with a $^\dagger$ are potential eclipsing white dwarf $+$ brown dwarf binaries based on PanSTARRS $(g-y)$ color and Gaia eDR3 parallax (see main text).}
  \label{tab:wdm_candidates_table}
\end{table*}

%% file: figures/wd_bd_cmd.tex
\begin{figure}
\includegraphics[scale=0.5]{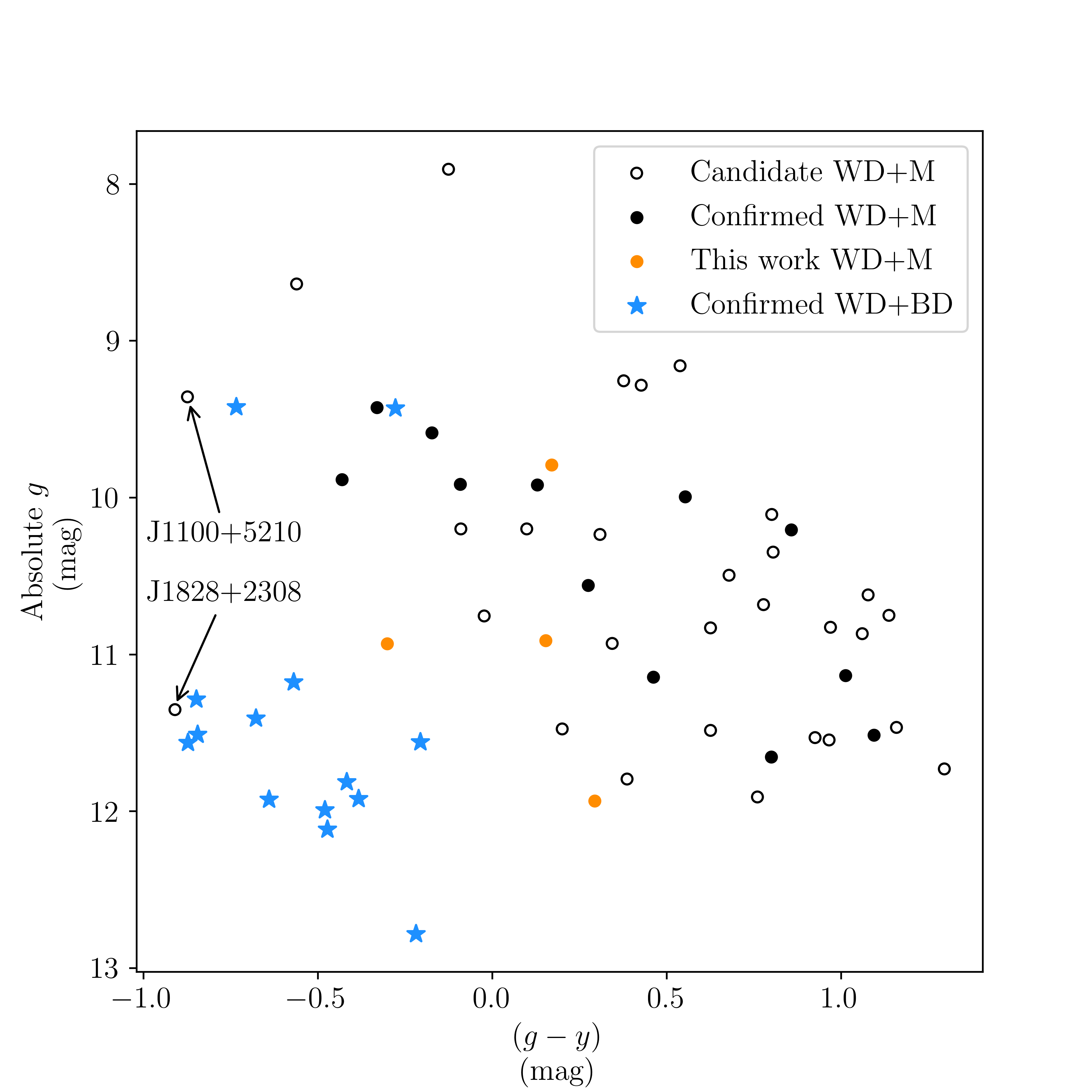}
\caption{Color-magnitude diagram using Gaia eDR3 parallax and Pan-STARRS $(g-y)$ color. Open/closed black circles represent the candidate/confirmed 41 WD$+$M binaries presented in Table \ref{tab:wdm_candidates_table}. Blue stars represent the 14 confirmed white dwarf $+$ brown dwarf binaries. We mark the locations of two new short-period, eclipsing, white dwarf $+$ brown dwarf candidates. The four new eclipsing WD$+$M binaries presented in this work are shown as filled orange circles.}
\label{fig:wd_bd_cmd}
\end{figure}

%% file: figures/wd_bd_lc.tex
  \begin{figure*}
    \hspace*{-1.35cm}
    \includegraphics[scale=0.45]{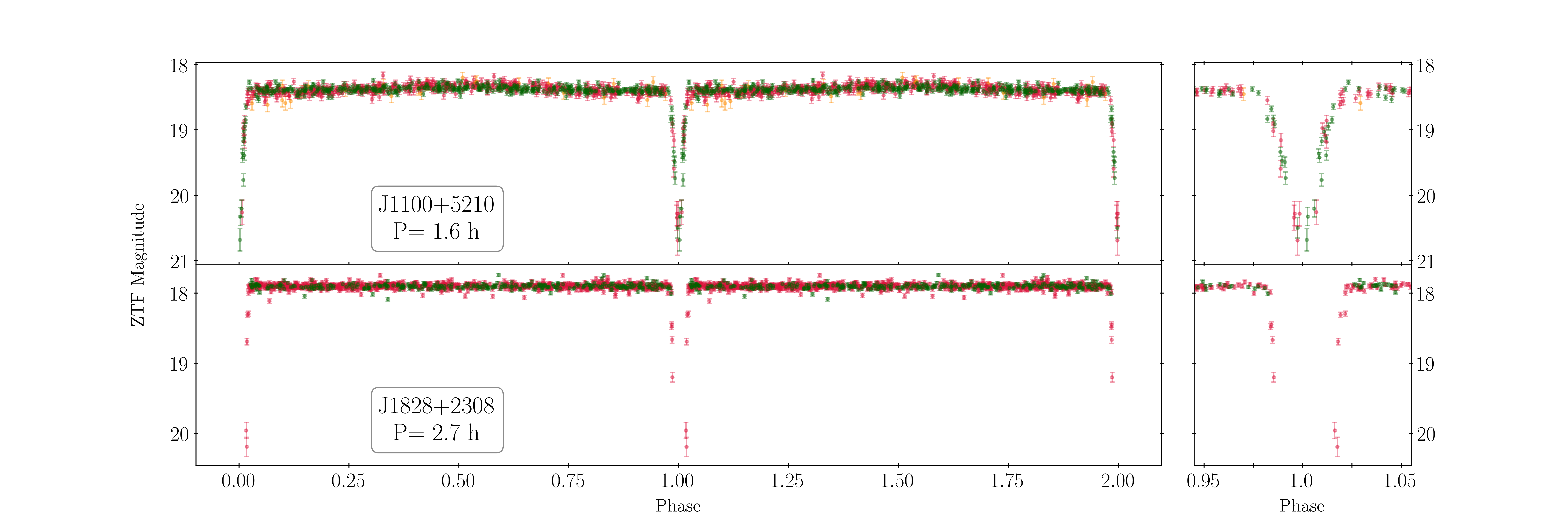}
    \caption{Public ZTF DR7 light curves for the two eclipsing WD$+$BD candidates identified in this work. Each light curve has been phase-folded to its most probable period, obtained through a box least squares period-finding algorithm \citep{kovacs2002}. Individual data points are colored based on which filter they were measured in: green points used the ZTF $g$-band, red points used the ZTF $r$-band, and orange points used the ZTF $i$-band.
    Data across all filters have been median combined to the median value of the ZTF $g$-band filter.}
    \label{fig:wd_bd_lc}
  \end{figure*}